\documentclass[12pt]{article}
\usepackage[dvips]{graphicx}
\usepackage{epsfig}
\usepackage{latexsym,amsmath,amsfonts,amssymb}
\usepackage{mathtools} 
\usepackage{slashed} 
\usepackage{blkarray}
\usepackage{cite}
\usepackage[all]{xy} 
\usepackage[makeroom]{cancel}
\usepackage[latin1]{inputenc}
\usepackage[american]{babel}
\usepackage[dvips]{graphicx}
\usepackage{bbm}
\usepackage{color}
\usepackage[unicode]{hyperref}
\def\im{Invent. Math.}

\def\a{\alpha}
\def\b{\beta}
\def\c{\gamma}
\def\d{\delta}
\def\f{\phi}               
\def\vf{\varphi}  \def\tvf{\tilde{\varphi}}
\def\vp{\varphi}
\def\g{\gamma}
\def\h{\eta}
\def\j{\psi}
\def\k{\kappa}                    
\def\l{\lambda}
\def\m{\mu}
\def\n{\nu}
\def\o{\omega}  \def\w{\omega}

\def\q{\theta}  \def\th{\theta}                  
\def\r{\rho}                                     
\def\s{\sigma}                                   
\def\t{\tau}
\def\u{\upsilon}
\def\x{\xi}
\def\z{\zeta}
\def\pt{\tilde{\varphi}}
\def\tt{\tilde{\theta}}
\def\lab{\label}
\def\6{\partial}
\def\wg{\wedge}
\def\bpsi{\bar{\psi}}
\def\bt{\bar{\theta}}
\def\bvf{\bar{\varphi}}

\DeclareMathOperator{\tr}{tr}

\newcommand{\be}{\begin{equation}}
\newcommand{\ee}{\end{equation}}
\newcommand{\beq}{\begin{equation}}
\newcommand{\eeq}{\end{equation}}
\newcommand{\bea}{\begin{eqnarray}}
\newcommand{\eea}{\end{eqnarray}}
\newcommand{\nn}{\nonumber \\}

\newcommand{\ba}{\begin{eqnarray}}
\newcommand{\ea}{\end{eqnarray}}

\newcommand{\beqs}{\begin{eqnarray}}
\newcommand{\eeqs}{\end{eqnarray}}
\newcommand{\bal}{\begin{aligned}}
\newcommand{\eal}{\end{aligned}}

\begin{document}
\baselineskip=15.5pt
\pagestyle{plain}
\setcounter{page}{1}


\def\del{{\partial}}
\def\vev#1{\left\langle #1 \right\rangle}
\def\cn{{\cal N}}
\def\co{{\cal O}}
\def\IC{{\mathbb C}}
\def\IR{{\mathbb R}}
\def\IZ{{\mathbb Z}}
\def\RP{{\bf RP}}
\def\CP{{\bf CP}}
\def\Poincare{{Poincar\'e }}
\def\tr{{\rm tr}}
\def\tp{{\tilde \Phi}}

\def\TL{\hfil$\displaystyle{##}$}
\def\TR{$\displaystyle{{}##}$\hfil}
\def\TC{\hfil$\displaystyle{##}$\hfil}
\def\TT{\hbox{##}}
\def\HLINE{\noalign{\vskip1\jot}\hline\noalign{\vskip1\jot}}
\def\seqalign#1#2{\vcenter{\openup1\jot
   \halign{\strut #1\cr #2 \cr}}}
\def\lbldef#1#2{\expandafter\gdef\csname #1\endcsname {#2}}
\def\eqn#1#2{\lbldef{#1}{(\ref{#1})}%
\begin{equation} #2 \label{#1} \end{equation}}
\def\eqalign#1{\vcenter{\openup1\jot
     \halign{\strut\span\TL & \span\TR\cr #1 \cr
    }}}
\def\eno#1{(\ref{#1})}
\def\href#1#2{#2}
\def\half{\frac{1}{2}}

\def\ads{{\it AdS}}
\def\adsp{{\it AdS}$_{p+2}$}
\def\cft{{\it CFT}}

\newcommand{\ber}{\begin{eqnarray}}
\newcommand{\eer}{\end{eqnarray}}

\newcommand{\beqar}{\begin{eqnarray}}
\newcommand{\cN}{{\cal N}}
\newcommand{\cO}{{\cal O}}
\newcommand{\cA}{{\cal A}}
\newcommand{\cT}{{\cal T}}
\newcommand{\cF}{{\cal F}}
\newcommand{\cC}{{\cal C}}
\newcommand{\cR}{{\cal R}}
\newcommand{\cW}{{\cal W}}
\newcommand{\eeqar}{\end{eqnarray}}
\newcommand{\tht}{\thteta}
\newcommand{\lm}{\lambda}\newcommand{\Lm}{\Lambda}
\newcommand{\R}{\mathbb{R}}


\newcommand{\nonu}{\nonumber}
\newcommand{\oh}{\displaystyle{\frac{1}{2}}}
\newcommand{\dsl}
   {\kern.06em\hbox{\raise.15ex\hbox{$/$}\kern-.56em\hbox{$\partial$}}}
\newcommand{\id}{i\!\!\not\!\partial}
\newcommand{\as}{\not\!\! A}
\newcommand{\ps}{\not\! p}
\newcommand{\ks}{\not\! k}
\newcommand{\D}{{\cal{D}}}
\newcommand{\dv}{d^2x}
\newcommand{\Z}{{\cal Z}}
\newcommand{\N}{{\cal N}}
\newcommand{\Dsl}{\not\!\! D}
\newcommand{\Bsl}{\not\!\! B}
\newcommand{\Psl}{\not\!\! P}
\newcommand{\eeqarr}{\end{eqnarray}}
\newcommand{\ZZ}{{\rm \kern 0.275em Z \kern -0.92em Z}\;}


\def\del{{\delta^{\hbox{\sevenrm B}}}} \def\ex{{\hbox{\rm e}}}
\def\azb{A_{\bar z}} \def\az{A_z} \def\bzb{B_{\bar z}} \def\bz{B_z}
\def\czb{C_{\bar z}} \def\cz{C_z} \def\dzb{D_{\bar z}} \def\dz{D_z}
\def\im{{\hbox{\rm Im}}} \def\mod{{\hbox{\rm mod}}} \def\tr{{\hbox{\rm Tr}}}
\def\ch{{\hbox{\rm ch}}} \def\imp{{\hbox{\sevenrm Im}}}
\def\trp{{\hbox{\sevenrm Tr}}} \def\vol{{\hbox{\rm vol}}}
\def\rl{\Lambda_{\hbox{\sevenrm R}}} \def\wl{\Lambda_{\hbox{\sevenrm W}}}
\def\fc{{\cal F}_{k+\cox}} \def\vev{vacuum expectation value}
\def\nodiv{\mid{\hbox{\hskip-7.8pt/}}}
\def\ie{{\em i.e.}}
\def\ie{\hbox{\it i.e.}}

\def\CC{{\mathchoice
{\rm C\mkern-8mu\vrule height1.45ex depth-.05ex
width.05em\mkern9mu\kern-.05em}
{\rm C\mkern-8mu\vrule height1.45ex depth-.05ex
width.05em\mkern9mu\kern-.05em}
{\rm C\mkern-8mu\vrule height1ex depth-.07ex
width.035em\mkern9mu\kern-.035em}
{\rm C\mkern-8mu\vrule height.65ex depth-.1ex
width.025em\mkern8mu\kern-.025em}}}

\def\RR{{\rm I\kern-1.6pt {\rm R}}}
\def\NN{{\rm I\!N}}
\def\ZZ{{\rm Z}\kern-3.8pt {\rm Z} \kern2pt}
\def\IB{\relax{\rm I\kern-.18em B}}
\def\ID{\relax{\rm I\kern-.18em D}}
\def\II{\relax{\rm I\kern-.18em I}}
\def\IP{\relax{\rm I\kern-.18em P}}
\newcommand{\CS}{{\scriptstyle {\rm CS}}}
\newcommand{\CSs}{{\scriptscriptstyle {\rm CS}}}
\newcommand{\rc}{\nonumber\\}
\newcommand{\bear}{\begin{eqnarray}}
\newcommand{\eear}{\end{eqnarray}}

\newcommand{\LL}{{\cal L}}

\def\mani{{\cal M}}
\def\calo{{\cal O}}
\def\calb{{\cal B}}
\def\calw{{\cal W}}
\def\calz{{\cal Z}}
\def\cald{{\cal D}}
\def\calc{{\cal C}}
\def\to{\rightarrow}
\def\ele{{\hbox{\sevenrm L}}}
\def\ere{{\hbox{\sevenrm R}}}
\def\zb{{\bar z}}
\def\wb{{\bar w}}
\def\nodiv{\mid{\hbox{\hskip-7.8pt/}}}
\def\menos{\hbox{\hskip-2.9pt}}
\def\dr{\dot R_}
\def\drr{\dot r_}
\def\ds{\dot s_}
\def\da{\dot A_}
\def\dga{\dot \gamma_}
\def\ga{\gamma_}
\def\dal{\dot\alpha_}
\def\al{\alpha_}
\def\cl{{closed}}
\def\cls{{closing}}
\def\vev{vacuum expectation value}
\def\tr{{\rm Tr}}
\def\to{\rightarrow}
\def\too{\longrightarrow}


\def\a{\alpha}
\def\b{\beta}
\def\c{\gamma}
\def\d{\delta}
\def\e{\epsilon}           
\def\F{\Phi}
\def\f{\phi}               
\def\vf{\varphi}  \def\tvf{\tilde{\varphi}}
\def\vp{\varphi}
\def\g{\gamma}
\def\h{\eta}
\def\j{\psi}
\def\k{\kappa}                    
\def\l{\lambda}
\def\m{\mu}
\def\n{\nu}
\def\o{\omega}  \def\w{\omega}
\def\q{\theta}  \def\th{\theta}                  
\def\r{\rho}                                     
\def\s{\sigma}                                   
\def\t{\tau}
\def\u{\upsilon}
\def\x{\xi}
\def\X{\Xi}
\def\z{\zeta}
\def\pt{\tilde{\varphi}}
\def\tt{\tilde{\theta}}
\def\lab{\label}
\def\6{\partial}
\def\wg{\wedge}
\def\atanh{{\rm arctanh}}
\def\bpsi{\bar{\psi}}
\def\bt{\bar{\theta}}
\def\bvf{\bar{\varphi}}

%

\newfont{\namefont}{cmr10}
\newfont{\addfont}{cmti7 scaled 1440}
\newfont{\boldmathfont}{cmbx10}
\newfont{\headfontb}{cmbx10 scaled 1728}
\newcommand{\re}{\,\mathbb{R}\mbox{e}\,}
\newcommand{\hyph}[1]{$#1$\nobreakdash-\hspace{0pt}}
\providecommand{\abs}[1]{\lvert#1\rvert}
\newcommand{\Nugual}[1]{$\mathcal{N}= #1 $}
\newcommand{\sub}[2]{#1_\text{#2}}
\newcommand{\partfrac}[2]{\frac{\partial #1}{\partial #2}}
\newcommand{\bsp}[1]{\begin{equation} \begin{split} #1 \end{split} \end{equation}}
\newcommand{\calF}{\mathcal{F}}
\newcommand{\calO}{\mathcal{O}}
\newcommand{\calM}{\mathcal{M}}
\newcommand{\calV}{\mathcal{V}}
\newcommand{\bbZ}{\mathbb{Z}}
\newcommand{\bbC}{\mathbb{C}}
\newcommand{\cK}{{\cal K}}
\newcommand{\dd}{\textrm{d}}
\newcommand{\DD}{\textrm{D}}

\newcommand{\Thq}{\Theta\left(\r-\r_q\right)}
\newcommand{\Dq}{\d\left(\r-\r_q\right)}
\newcommand{\kten}{\kappa^2_{\left(10\right)}}
\newcommand{\pbi}[1]{\imath^*\left(#1\right)}
\newcommand{\ho}{\hat{\omega}}
\newcommand{\tth}{\tilde{\th}}
\newcommand{\tf}{\tilde{\f}}
\newcommand{\tj}{\tilde{\j}}
\newcommand{\tw}{\tilde{\omega}}
\newcommand{\tz}{\tilde{z}}
\newcommand{\prj}[2]{(\partial_r{#1})(\partial_{\j}{#2})-(\partial_r{#2})(\partial_{\j}{#1})}
\def\atanh{{\rm arctanh}}
\def\sech{{\rm sech}}
\def\csch{{\rm csch}}
\allowdisplaybreaks[1]

\def\red{\textcolor[rgb]{0.98,0.00,0.00}}
\def\blue{\textcolor[rgb]{0.00,0.00,0.98}}

\numberwithin{equation}{section}

\newcommand{\Tr}{\mbox{Tr}}    


%
\renewcommand{\theequation}{{\rm\thesection.\arabic{equation}}}

\begin{titlepage}
\vskip 3mm
\begin{flushright}
APCTP Pre2018-003\\
IPM/P-2018/010
\end{flushright}
\begin{center}
   \baselineskip=16pt
  \centerline{ {\LARGE \bf Yang-Baxter Deformations Beyond Coset Spaces }}
  \vskip 0.3cm
  \centerline{\Large \bf (a slick way to do TsT)}
   \vskip 6mm
\centerline{ {\large \bf    I. Bakhmatov$^{a, b}$, E. \'O Colg\'ain$^{a, c}$, M. M. Sheikh-Jabbari$^d$, H. Yavartanoo$^e$}}
       \vskip .6cm
             \begin{small}
               \textit{$^a$ Asia Pacific Center for Theoretical Physics, Postech, Pohang 37673, Korea}
               
               \vspace{2mm} 
               
               \textit{$^b$ Institute of Physics, Kazan Federal University, Kremlevskaya 16a, 420111, Kazan, Russia }
               
               \vspace{2mm}
               
               \textit{$^c$ Department of Physics, Postech, Pohang 37673, Korea}
               
               \vspace{2mm}
               
               \textit{$^d$ School of Physics, Institute for Research in Fundamental Sciences (IPM), \\ P.O.Box 19395-5531, Tehran, Iran}
               
               \vspace{2mm} 
               
               \textit{$^e$ State Key Laboratory of Theoretical Physics, Institute of Theoretical Physics,\\
Chinese Academy of Sciences, Beijing 100190, China}
  
             \end{small}
\end{center}
\vskip 3mm
\begin{center} \textbf{Abstract}\end{center} \begin{quote}

Yang-Baxter string sigma-models provide a systematic way to deform coset geometries, such as $AdS_p \times S^p$, while retaining the $\sigma$-model integrability. It has been shown that the Yang-Baxter deformation in target space is simply an open-closed string map that can be defined for any geometry, not just coset spaces. Given a geometry with an isometry group and a bivector that is assumed to be a linear combination of antisymmetric products of Killing vectors, we show the equations of motion of (generalized) supergravity reduce to the Classical Yang-Baxter Equation associated with the isometry group, proving the statement made in \cite{Bakhmatov:2017joy}. These results bring us closer to the proof of  the ``YB solution generating technique'' for (generalized) supergravity advertised in \cite{Bakhmatov:2017joy} and in particular provide an economical way to perform TsT transformations.

\end{quote} \vfill

\end{titlepage}


\section{Introduction}

Klimcik's pioneering work on integrable deformations of $\sigma$-models \cite{Klimcik:2002zj, Klimcik:2008eq} paved the way for their application to string $\sigma$-models and AdS/CFT geometries \cite{Delduc:2013qra,Kawaguchi:2014qwa}. Thanks to this breakthrough, we now understand noncommutative \cite{Hashimoto:1999ut, Maldacena:1999mh, Alishahiha:1999ci} and marginal deformations \cite{Lunin:2005jy,Frolov:2005dj} of AdS/CFT geometries in a new light: they are part of a larger family of integrable deformations of $AdS_p \times S^p$ geometries, where the deformation is given by an $r$-matrix solution to the Classical Yang-Baxter Equation (CYBE). One exciting outcome of this research line has been the observation that the deformed geometries  may not be consistent string theory backgrounds in the usual sense; they are not solutions of usual supergravity. Nonetheless, it has been noted that they are solutions to the generalized supergravity equations of motion \cite{Arutyunov:2015mqj, Wulff:2016tju}, which differ from usual supergravity through an additional Killing vector $I$. Furthermore, exploring the connection to TsT transformations \cite{Matsumoto:2014nra, Matsumoto:2014gwa}, established in \cite{Osten:2016dvf}, it was conjectured \cite{Hoare:2016wsk}, and later proved \cite{Borsato:2016pas, Borsato:2017qsx} that homogeneous Yang-Baxter deformations are equivalent to non-Abelian duality transformations \footnote{See \cite{Arutyunov:2013ega,Kawaguchi:2014fca,Hoare:2014pna,Arutynov:2014ota,Khouchen:2014kaa,Ahn:2014aqa,Arutyunov:2014cra,Arutyunov:2014cda, Banerjee:2014bca,Kameyama:2014vma, Kameyama:2014via,Lunin:2014tsa, Hoare:2014oua, Matsumoto:2014ubv,Engelund:2014pla,Ahn:2014iia,Panigrahi:2014sia,Bai:2014pya,Matsumoto:2015jja,Bozhilov:2015kya,Matsumoto:2015uja,Banerjee:2015nha, vanTongeren:2015soa, Vicedo:2015pna, Hoare:2015gda,Khouchen:2015jfa,vanTongeren:2015uha,Sfetsos:2015nya, Arutyunov:2015qva, Hoare:2015wia, Klimcik:2015gba,Kameyama:2015ufa, Borowiec:2015wua,Demulder:2016mja,Kyono:2016jqy,Hoare:2016ibq,Hoare:2016hwh,Delduc:2016ihq,Klimcik:2016rov,Orlando:2016qqu,Banerjee:2016xbb,Arutyunov:2016ysi,vanTongeren:2016eeb,Hoare:2016wca,Ahn:2016egk,Roychowdhury:2016bsv,Sakamoto:2016ppx,Delduc:2017brb,Kyono:2017jtc,Roychowdhury:2017oqd,Appadu:2017bnv,Klimcik:2017ken,Roychowdhury:2017vdo,Hernandez:2017raj,Delduc:2017fib,Klabbers:2017vtw,Hoare:2017ukq,Demulder:2017zhz,Banerjee:2017mpe,Barik:2018haz,Lust:2018jsx} for a sample of related developments.}.

A subsequent simpler proposal surfaced in \cite{Araujo:2017jkb,Araujo:2017jap,Araujo:2017enj}, where it was shown that the closed-open string map, which is an extension of the map between open string and closed string frames initially introduced by Seiberg \& Witten \cite{Seiberg:1999vs} (see also \cite{Duff:1989tf}), undoes all known Yang-Baxter deformations. The obvious implication of this finding is that Yang-Baxter $\sigma$-models are really the open-closed string map in disguise. While this observation may have various bearings for open or closed string theories on these backgrounds, in this work we focus on the backgrounds themselves and explore a potentially powerful solution generating technique for (generalized) supergravity. For backgrounds where the Killing vector $I$ is not sourced, our methods constitute a supergravity solution generating technique. 

More concretely, given a spacetime metric $G_{\mu \nu}$ with an isometry group and $\Theta^{\mu \nu}$ an $r$-matrix solution to the CYBE (in Killing vector representation) of the associated Lie algebra, the deformed geometry, consisting of a metric $g_{\mu \nu}$ and NSNS two-form $B_{\mu \nu}$, is defined through the map:
\be
\label{map}
 (G^{-1} + \Theta)^{-1} = (g +B),
\ee
where it is worth noting that $\Theta = 0$ implies $B= 0$. This transformation (\ref{map}), which holds in the $\sigma$-model target space, is the essence of Yang-Baxter $\sigma$-models. There is no condition that the original geometry be  a coset and no moving parts: this is simply a high-school level matrix inversion. Moreover, as it stands (\ref{map}) is \textit{a priori} valid for any spacetime metric $G_{\mu \nu}$. Building on this observation, a general prescription for transforming the dilaton, RR sector, as well as introducing a Killing vector $I$, was presented in \cite{Bakhmatov:2017joy}, where the method was applied to explicit coset and non-coset geometries alike \footnote{In a series of papers \cite{Sakamoto:2017cpu, Fernandez-Melgarejo:2017oyu, Sakamoto:2018krs}, the same map has been embedded in DFT, where $\Theta = \beta$, but with a continued focus on coset geometries. Our prescriptions for transforming fields, including the RR sector, are expected to agree.}. In this paper, we move beyond examples and move towards a general proof of the statements in \cite{Bakhmatov:2017joy}.

Before proceeding, let us briefly take stock. The map (\ref{map}) may look like a straightforward generalisation of the Yang-Baxter $\sigma$-model to general spacetimes, but there is a palpable difference in philosophy. In the traditional Yang-Baxter $\sigma$-model narrative of Klimcik \cite{Klimcik:2002zj, Klimcik:2008eq}, the $r$-matrix solution to the CYBE is an input and this is the magic ingredient that guarantees integrability of the deformed coset $\sigma$-model. Here, we relax this input and adopt the milder assumption that $\Theta$ is bi-Killing. Recalling that CYBE is an algebraic equation on a given Lie algebra, this bi-Killing structure is well justified. To ``geometrise'' the CYBE, it is hence natural to consider Killing vectors and the isometry algebra of a given geometry. The coefficients of the bi-Killing antisymmetric bivector $\Theta$ are then a constant skew-symmetric matrix  $r$. As noted for explicit examples in \cite{Bakhmatov:2017joy}, the dynamical equations of motion (EOMs) of generalized supergravity then reduce to the purely algebraic CYBE on $r$. Thus, the CYBE becomes the output. Moreover, the connection to integrability is severed, since it is clear that even for non-integrable geometries \footnote{As remarked earlier for the geometry $T^{1,1}$ \cite{Crichigno:2014ipa}} the map (\ref{map}) exists. Our observation ultimately means that supergravity can be exploited to classify solutions to the CYBE, providing a striking application of physics to a mathematics problem. Conversely, our analysis provides a solution generating technique for (generalized) supergravity. We start from any solution, construct a $\Theta$ from solutions to the CYBE associated with the isometries of the background, then use this data and the above map (\ref{map}) to construct the deformed background, which is a solution to generalized supergravity. This method as we demonstrate later provides a more economical way to perform TsT transformations.

As mentioned, the purpose of this current manuscript is to substantiate the claims of \cite{Bakhmatov:2017joy} by moving beyond examples to generic spacetimes. When working with solution generating techniques in supergravity, it is the case that once one nails the NS sector, the transformation of the RR sector can be pieced together \footnote{For T-duality and Yang-Baxter $\sigma$-models, it is well-known that the frame rotation manifests itself in a Lorentz transformation on the flux bispinor \cite{Hassan:1999bv, Kelekci:2014ima, Borsato:2016ose,Sakamoto:2018krs}.}. For this reason, we focus purely on the NS sector of generalized supergravity, while extension to the full generalized supergravity with inclusion of the RR sector is just an added technicality. Furthermore, since the map is only defined for geometries with vanishing NSNS two-form, we are forced to restrict ourselves to geometries that are supported by a scalar dilaton $\Phi$, which guarantees that they are not only Ricci-flat but can be curved. While it is easy to invert (\ref{map}) for explicit solutions, such as $AdS_2 \times S^2$ and the Schwarzschild black hole \cite{Bakhmatov:2017joy}, for arbitrary $G$ and $\Theta$ extracting $g$ and $B$, so that one can check the EOMs, is challenging.

To overcome this difficulty, we work perturbatively in the deformation parameter $\Theta$ about an arbitrary background $G$, which may be supported by a scalar dilaton $\Phi$. This allows us to expand $g$ and $B$ in $\Theta$ and substitute the expressions directly into the EOMs of generalized supergravity \cite{Arutyunov:2015mqj}.  However, it turns out that little progress can be made for generic $\Theta$, so we are forced to also assume that it is bi-Killing,
\be\label{bi-Killing}
\Theta^{\mu \nu} =  r^{ij} K^{\mu}_i K^{\nu}_j,
\ee
where $K_i$ denote Killing vectors of the background and the constant coefficients are skew-symmetric, $r^{ij} = - r^{ji}$. Doing so, we arrive at a number of results, which we have checked to third order in $\Theta$. At first order, we are able to prove that the Killing vector $I$ of generalized supergravity is the divergence of the bivector,
\be
\label{I_theta}
I^{\mu} = \nabla_{\nu} \Theta^{\nu \mu},
\ee
thus providing a proof of a relation identified earlier in \cite{Araujo:2017jkb,Araujo:2017jap}. Since $\Theta$ is bi-Killing, it should be noted that $I$ is Killing by construction. Previously this relation was motivated by the $\Lambda$-gauge symmetry of the NSNS two-form, $B \rightarrow B + \dd \Lambda$, where $\Lambda$ is an arbitrary one-form \cite{Araujo:2017enj}. At second order, we confirm that the dilaton and Einstein equation reduce to the CYBE. Imposing the CYBE we find that the third order equations are trivially satisfied, which is consistent with the claim of \cite{Bakhmatov:2017joy} that the EOMs are equivalent to the CYBE once $\Theta$ is bi-Killing. For explicit solutions, it is possible to go further and check our claim to all orders.

The structure of this paper runs as follows. In section \ref{sec:prel} we introduce the bi-Killing structure of $\Theta$ and show that the Jacobi identity for $\Theta$ is equivalent to the homogeneous CYBE. We also explain how the dilaton transforms. In section \ref{sec:pert} we study the map (\ref{map}) perturbatively to third order in $\Theta$, in the process proving (\ref{I_theta}) and demonstrating that the CYBE emerges from the Einstein and dilaton EOM at second order, as well as the $B$-field EOM at third order. In section \ref{sec:ex}, we provide deformations of flat spacetime, Bianchi spacetimes and provide an example that includes the RR sector. In particular, we confirm that the  Lunin-Maldacena-Frolov geometries \cite{Lunin:2005jy, Frolov:2005dj} can be easily recovered using the methods outlined in \cite{Bakhmatov:2017joy}. Thus, Yang-Baxter deformations provide a smart way to perform TsT transformations and there is no need to resort to T-duality transformations. For sadomasochists, gory details can be found in the appendix.

\section{Preliminaries}
\label{sec:prel}
In this section, we provide a setting for later calculations. We start with a description of the bi-Killing structure of the bivector $\Theta$ in the open-closed string map (\ref{map}). We recall that we are considering generic spacetime metrics $G_{\mu \nu}$ with an isometry group. From the Killing vectors $K_i$, one can construct an antisymmetric product of Killing vectors, 
\be
\label{theta}
\Theta^{\mu \nu} = \frac{r^{ij}}{2} ( K^{\mu }_{i} K^{\nu}_j - K^{\nu}_i K^{\mu}_j ) = r^{ij} K^{\mu}_{i} K^{\nu}_j,  
\ee
where $r^{ij}$ is a skew-symmetric matrix, $r^{ij} = - r^{ji}$, with constant coefficients. The above ansatz is motivated by the fact that we are exploring possible connections between the CYBE over the isometry algebra of a given solution and a class of deformations.  Being Killing vectors associated to an isometry group, $K_i$ satisfy the commutation relation
\be
[ K_i, K_j] = c_{ij}^{~~k} K_k, 
\ee
where $c_{ij}^{~~k}$ denote the structure constants. One may recast this relation in terms of components as, 
\be
\label{comm}
K_i^{\rho} \nabla_{\rho} K_j^{\gamma} - K_j^{\rho} \nabla_{\rho} K_{i}^{\gamma} = c_{ij}^{~~k} K_{k}^{\gamma}, 
\ee
where it makes no difference if one replaces the covariant derivatives with  partial derivatives, since the Christoffel symbols cancel. 

We recall that the map (\ref{map}) appears in the string theory literature in the context of noncommutativity in string theory \cite{Seiberg:1999vs}, where $\Theta$ is the noncommutativity (NC) parameter. In the open string setting, the endpoints of the open string parametrised by $X^\mu$ coordinates satisfy a commutation relation
\be
[X^\mu,X^\nu]=i\Theta^{\mu\nu}(X).
\ee
For the above algebra to be consistent, the NC parameter should satisfy the  Jacobi identity 
\be
\label{Jacobi} 
\Theta^{\alpha \rho} \nabla_{\rho} \Theta^{\beta \gamma} + \Theta^{\gamma \rho} \nabla_{\rho} \Theta^{\alpha \beta} + \Theta^{\beta \rho} \nabla_{\rho} \Theta^{\gamma \alpha} = 0. 
\ee

Using the bi-Killing structure of $\Theta$, it is an easy exercise to show that the Jacobi identity is equivalent to the CYBE. First we consider 
\be
\Theta^{\alpha \rho} \nabla_{\rho} \Theta^{\beta \gamma} = r^{ij} r^{kl} \left( K_{i}^{\alpha} K_{l}^{\gamma} K^{\rho}_{j} \nabla_{\rho} K_k^{\beta} + K_{i}^{\alpha} K_k^{\beta} K^{\rho}_j \nabla_{\rho} K_{l}^{\gamma} \right),  
\ee
before antisymmetrising, 
\bea\label{Jacobi-Theta}
\Theta^{[\alpha \rho} \nabla_{\rho} \Theta^{\beta \gamma]} 
&=& K^{\alpha}_i  K^{\beta}_j K^{\gamma}_{k} ( c_{l_1 l_2}^{~~~i} r^{j l_1} r^{k l_2} + c_{l_1 l_2}^{~~~k} r^{ i l_1} r^{ j l_2} + c_{l_1 l_2}^{~~~j} r^{ k l_1} r^{ i l_2} ) = 0.   
\eea
Modulo the Killing vectors, the RHS is the homogeneous CYBE\footnote{We note that since \eqref{Jacobi-Theta} should hold for all points on spacetime, then it can be satisfied only if the constant, spacetime independent piece vanishes.}
\be
\label{CYBE} c_{l_1 l_2}^{~~~[i} r^{j l_1} r^{k] l_2} = 0.
\ee
See \cite{Sakamoto:2017cpu, Fernandez-Melgarejo:2017oyu, Sakamoto:2018krs} for a similarly explicit derivation of the relation between the CYBE and the vanishing of R-flux, essentially the Jacobi identity above.  We also note that the left hand side of \eqref{Jacobi} is also known as the Schouten bracket in the context of double field theory and general $O(d,d)$ string theory compactifications \cite{Sakamoto:2017cpu}. 

At this point, it is an opportune time to recall that $r$-matrix solutions to the CYBE take the form
\be
r = \frac{1}{2} r^{ij} T_i \wedge T_j, 
\ee
where $T_i$ are elements of the Lie algebra, $T_i \in \frak{g}$. It should now be clear that the bi-Killing structure of $\Theta$ mimics the $r$-matrix. In other words,  one can assume that $\Theta$ is the $r$-matrix written in the basis of Killing vectors.  This relation has been observed for all Yang-Baxter deformations, even for $r$-matrix solutions to the modified CYBE \cite{Araujo:2017jap} (see Appendix B).  We note that for $G/H$ coset spaces the Killing vectors are basically the same as the generators of $G$ which also provide a complete basis for expanding any tensor. In this sense, the bi-Killing structure allows for $\Theta$ to have all possible components. Of course, the spacetime dependence of $\Theta$ is still not fixed by the bi-Killing assumption. 

At this stage, we have introduced the bivector $\Theta$, which plays a central role in our map (\ref{map}), and explained its bi-Killing structure. We have further demonstrated a connection between the interpretation of $\Theta$ as an NC parameter, which is required to satisfy the Jacobi identity, and its role as an $r$-matrix solution to the CYBE. We recall that NC deformations of field theories are intimately connected to Drinfeld twists \cite{Drinfeld} of Lie algebras, where the twist element is precisely an $r$-matrix solution to the CYBE \cite{Chaichian:2004za, Chaichian:2004yh, Lukierski:2005fc}.\footnote{We thank Anca Tureanu for a discussion on this point.} For this reason, it is expected that the Jacobi identity is the CYBE {when the bivector $\Theta$ is bi-Killing. Of course, it is more careful to state that the CYBE implies the Jacobi identity since there may be solutions to the Jacobi identity that are not bi-Killing.}

Moving on, we will now address the relation between the Killing vector $I$ of generalized supergravity \cite{Arutyunov:2015mqj} and the bivector $\Theta$ (\ref{I_theta}). In \cite{Araujo:2017jap} it was checked that this relation holds for a large class of solutions to generalized supergravity and we will prove it is an outcome of the generalized supergravity EOMs in the next section. Here, using the bi-Killing structure, we motivate this relation in a simple way. A short calculation reveals that 
\be
I^{\mu} = r^{ij} \nabla_{\nu} ( K^{\nu}_i K^{\mu}_j ) = r^{ij} K^{\nu}_i \nabla_{\nu} K^{\mu}_j  = \frac{1}{2} r^{ij} c_{ij}^{~~k} K^{\mu}_k, 
\ee
where we have used the fact that $r^{ij}$ is antisymmetric and the commutation relation (\ref{comm}). Therefore, by construction $I$ is a linear combination of Killing vectors with constant coefficients and is hence guaranteed to be Killing. As a further check, we note that when $I = 0$, so that the solution corresponds to a solution of usual supergravity, we recover the unimodularity condition of \cite{Borsato:2016ose}, 
\be
r^{ij} [ T_i, T_j] = 0 \quad \Rightarrow \quad r^{ij} c_{ij}^{~~k} = 0. 
\ee
In summary, given the fact that the relation (\ref{I_theta}) holds for a large number of explicit solutions \cite{Araujo:2017jap}, it can be explained for D-brane geometries \cite{Araujo:2017enj} and that it recovers a result in the independent literature \cite{Borsato:2016ose}, this should put any doubts about the validity of (\ref{I_theta}) to rest. That being said, we have yet to identify the Killing vector $I$ with the Killing vector appearing in the EOMs of generalized supergravity. This we will do in the next section. 

Before proceeding to the next section, where we will study the EOMs of generalized supergravity, it is important to address the transformation of the dilaton. At this stage, given the original metric $G$, our map (\ref{map}) and (\ref{I_theta}), the deformed metric $g$, NSNS two-form $B$ and Killing vector $I$ are completely determined in terms of $\Theta$. As proposed originally in \cite{Araujo:2017enj}, the usual T-duality density $e^{-2 \phi} \sqrt{g}$ (valid for both Abelian and non-Abelian T-duality) is invariant. With this assumption, given the metric $G$ and scalar $\Phi$, the transformed dilaton $\phi$ is 
\be
\label{dilaton}
e^{-2 \phi} \sqrt{g} = e^{-2 \Phi} \sqrt{G} \quad \Rightarrow \quad \phi = \Phi + \frac{1}{4} \log \left(\frac{g}{G}\right). 
\ee
This completes our treatment of the NS sector:
\begin{align}
\text{original solution:}\ G_{\mu\nu}, \Theta^{\mu\nu}, \Phi;&\qquad \text{deformed solution:}\ g_{\mu\nu}, B_{\mu\nu}, \phi,\\
g_{\mu\nu}=(G^{-1}- \Theta\cdot G\cdot \Theta)^{-1}_{\mu\nu}&,\qquad 
B_{\mu\nu}=-(G^{-1}-\Theta)^{-1} \cdot \Theta \cdot (G^{-1}+\Theta)^{-1}\cr
\phi&=\Phi-\frac12 \ln \det(1+G\cdot\Theta). \label{open-closed-map}
\end{align}
In the above $\cdot$ denotes matrix multiplication and $G$ and $\Theta$ are to be viewed as two matrices. The indices on $\Theta$ are lowered and raised by the metric $G$.
 We remind the reader that a complete prescription including the RR sector can be found in \cite{Bakhmatov:2017joy}.

\section{Perturbative analysis}
\label{sec:pert} 
In this section, we will extract the CYBE from the EOMs of generalized supergravity. As stated earlier, we restrict our attention to the NS sector on the basis that repeating the calculations for the RR sector will not offer new insights. Indeed, since we are working perturbatively, yet ultimately interested in exact solutions, we will fall short of our goal of establishing the map (\ref{map}) and dilaton transformation (\ref{dilaton}) as a solution generating technique. Instead, we will expand in $\Theta$ around a generic background and enumerate the conditions that should hold through third order in $\Theta$ so that a solution can exist. We will see that all conditions, including the CYBE, follow once one assumes that $\Theta$ is bi-Killing. 

As we have seen, the bi-Killing vector $\Theta$ is essentially the $r$-matrix. Since the CYBE is quadratic in components of the $r$-matrix, it is reasonable to expect that the CYBE emerges from the EOMs of generalized supergravity at second order in $\Theta$. For this reason, in this section, we expand our map (\ref{map}) to second order in $\Theta$ in the EOMs. At leading order, we identify conditions that are satisfied once $\Theta$ is bi-Killing and $I$ is a Killing vector, a fact that is guaranteed by the relation (\ref{I_theta}). At second order, we find from the $B$-field EOM that the Lie derivative of $\Theta$ with respect to $I$ must vanish, $\mathcal{L}_I \Theta = 0$, while from the Einstein and dilaton EOMs we recover the CYBE. Details of the calculations can be found in the appendix. 

\subsection{Review of generalized supergravity} 
Let us begin by recalling the EOMs of generalized supergravity \cite{Arutyunov:2015mqj},
 {
\bea
\label{B_eom}
\frac{1}{2} \hat \nabla^{\rho} H_{\rho \mu \nu} &=& X^{\rho} H_{\rho \mu \nu} + \hat\nabla_{\mu} X_{\nu} - \hat\nabla_{\nu} X_{\mu}, \\
\label{dilaton_EOM} \frac1{12}H^2 &=& 2 X_{\mu} X^{\mu}-  \hat\nabla_{\mu} X^{\mu}, \\
\label{Einstein}
\hat R_{\mu \nu} &=& \frac{1}{4} H_{\mu \rho \sigma} H_{\nu}^{~ \rho \sigma} - \hat\nabla_{\mu } X_{\nu} - \hat\nabla_{\nu} X_{\mu}, 
\eea
where $\hat \nabla$ and $\hat R_{\mu\nu}$ denote the covariant derivative and curvature of the deformed solution $g_{\mu \nu}$, we have used the trace of the Einstein equation to eliminate $\hat{R}$ in (\ref{dilaton_EOM}), and we have defined the one-form, 
\be\label{X-I}
X_{\mu} = \partial_{\mu} \phi + ( g_{\nu \mu} + B_{\nu \mu} ) I^{\nu}. 
\ee
{Throughout the remainder of this work, we will refer to the equations (\ref{B_eom}), (\ref{dilaton_EOM}) and (\ref{Einstein}) as the NSNS two-form ($B$-field) EOM, the dilaton EOM and the Einstein equation, respectively.} To derive these expressions \cite{Arutyunov:2015mqj} it has been assumed that $I$ is a Killing vector. Here, one can drop that assumption as this condition appears from the EOMs at leading order, thus providing a further consistency condition on the work presented in \cite{Arutyunov:2015mqj}. In other words, it is enough to assume the above equations. It is worth noting also that setting $I=0$, we recover the EOMs of usual supergravity.

We remark that the generalized gravity EOMs, similarly to the supergravity EOMs, are closely related to the string theory $\sigma$-model.  One may start from a $\sigma$-model obtained from a generic
 non-Abelian T-duality over a usual consistent string worldsheet theory. As a result of non-Abelian T-duality, the worldsheet anomaly cancelation does not lead to supergravity equation, but rather the generalized supergravity EOM \cite{Elitzur:1994ri} (see also \cite{Hull-Tonwsend} for earlier work and \cite{Arutyunov:2015mqj, Borsato:2017qsx, Arutyunov:2015qva} for addition of the RR-fields).  From the perspective of the $\sigma$-model, the Killing vector $I$ appears to be the trace of the structure constants of the non-semisimple group on which one T-dualises \cite{Hong:2018tlp}. In contrast to the original treatment of generalized supergravity \cite{Arutyunov:2015mqj}, where an explicit solution and T-duality on a non-isometric direction were used to motivate the EOMs, or \cite{Wulff:2016tju} where $\kappa$-symmetry is assumed, the derivation \cite{Hong:2018tlp} from the T-dual $\sigma$-model of \cite{Elitzur:1994ri} is purely bosonic and does not assume fermions \footnote{The analysis presented in \cite{Hong:2018tlp} is restricted to the NS sector.}. 

\paragraph{Check of consistency of the EOMs.} Regardless of their $\sigma$-model roots, one can ask if the generalized supergravity EOMs provide a consistent set of differential equations. For the set of equations \eqref{B_eom}, \eqref{dilaton_EOM}, \eqref{Einstein}, this amounts to checking if the Bianchi identity $\hat\nabla^\mu (\hat R_{\mu\nu}-\frac12 \hat R g_{\mu\nu})=0$ holds for any on-shell configuration. As the detailed calculations of the appendix demonstrates, straightforward but tedious analysis,  reveals that this identity is satisfied iff the one-form $X_{\mu}$ has the form \eqref{X-I} for an arbitrary Killing one-form field $I_\mu$. That is, \eqref{X-I} is also an outcome of the set of generalized supergravity EOMs and need not be put in by hand.  We also comment that while this consistency check is the necessary condition for the EOMs to come from a diffeomorphism invariant action, it is not sufficient; generalized supergravity is described by its EOMs and it is not known whether this theory has an action. 

\subsection{$\Theta$ expansion}

What we will do in this section is solve the EOMs by a perturbative expansion in powers of $\Theta$ around a given solution at $\Theta=0$. This latter is given by background metric $G_{\mu \nu}$ and dilaton $\Phi$. We start by expanding \eqref{open-closed-map} 
\bea\label{g-B-phi-expanded}
g_{\mu \nu} &=& G_{\mu \nu} + \Theta_{\mu}^{\ \ \alpha}\Theta_{\alpha\nu} +{\cal O}(\Theta^4), \nn
B_{\mu \nu} &=& - \Theta_{\mu \nu}  - \Theta_{\mu\alpha} \Theta^{\alpha\beta}\Theta_{\beta\nu} +{\cal O}(\Theta^5), \\
 \phi &=& \Phi + \frac{1}{4} \Theta_{\rho \sigma}\Theta^{\rho \sigma}+ {\cal O}(\Theta^4),\nonumber
\eea
where  all indices are raised and lowered with respect to the background metric $G_{\mu \nu}$. 

\paragraph{Zeroth order:} At this order the $B$-field equation is trivial and the other two equations read as
\be
R_{\mu\nu}+2\nabla_\mu\nabla_\nu\Phi=0,\qquad \nabla^2\Phi-2(\nabla\Phi)^2=0, 
\ee
where the curvature is computed using background metric $G_{\mu\nu}$.

\paragraph{First order:} At first order the dilaton and Einstein equations \eqref{dilaton_EOM} and \eqref{Einstein}, respectively yield
\be
I^\mu\nabla_\mu\Phi=0,\qquad \nabla_\mu I_\nu+\nabla_\nu I_\mu=0
\ee
which just confirm $I$ as a Killing vector for the background solution, specified by $G_{\mu\nu}, \Phi$.

The NSNS two-form EOM \eqref{B_eom} using the bi-Killing structure of $\Theta$ \eqref{bi-Killing}, after straightforward algebra and using Killing vector identities (see appendix for details), yields
\be
\nabla_\alpha(\nabla_\mu \Theta^{\mu\nu}-I^\nu)=0 \ \Longrightarrow \ I^\mu=\nabla_\nu\Theta^{\nu\mu}+const.
\ee
The constant part may be dropped using the fact that we want $I=0$ at zeroth order when $\Theta=0$. We hence recover \eqref{I_theta} as a consequence of the first order EOMs.

\paragraph{Second order:} The NSNS two-form EOM, once we use the first order results, takes a very simple form:
\be
\mathcal{L}_{I} \Theta = \dd i_{I} \Theta + i_{I} \dd \Theta = 0, 
\ee
which essentially tells us that $I$ is not only a Killing vector of the original geometry but also remains Killing in the deformed geometry. 

We next consider the dilaton and Einstein equations at second order.  To work these out, one should note that the covariant derivatives appearing in the EOMs are with respect to the metric $g_{\mu\nu}$ and hence one should expand the Christoffel symbols too,  {
\be
\hat \nabla_\mu X_\nu=\nabla_\mu X_\nu-\frac12G^{\rho\alpha}(\hat\nabla_\mu\Theta^2_{\alpha\nu}+\hat\nabla_\nu\Theta^2_{\alpha\mu}-\hat\nabla_\alpha\Theta^2_{\mu\nu}    ) X_\rho+\cdots,
\ee
where $\nabla_\mu$ } denotes covariant derivative with respect to the metric $G_{\mu\nu}$ and $\cdots$ stand for higher powers of $\Theta$. The Riemann curvature then receives even power corrections due to the correction to the Christoffel connection. The $H^2$ terms in the EOM also contribute to the second and all even powers. 

After lengthy calculations, the dilaton equation of motion takes the form
\bea
K^{\alpha}_i K^{\beta}_k \nabla_{\alpha} K_{\beta m} \left( c_{l_1 l_2}^{~~~ m} r^{i l_1} r^{k l_2} + c_{l_1 l_2}^{~~~ k} r^{m l_1} r^{i l_2} + c_{l_1 l_2}^{~~~ i} r^{k l_1} r^{m l_2} \right)&+& \nn\left(\Theta^{\beta \gamma} \Theta^{\alpha \lambda} + \Theta^{\alpha \beta} \Theta^{\gamma \lambda} + \Theta^{ \gamma \alpha} \Theta^{\beta \lambda} \right) R_{\beta \gamma \alpha \lambda} &=& 0.
\eea
The second line vanishes due to  the Bianchi $R_{[\alpha \beta \gamma] \lambda} = 0$ and the first line yields  the CYBE \eqref{CYBE}.  The Einstein equation can also be massaged and brought to the form \bea
 \frac{1}{2}  ( {\nabla}_{\rho} K_{ i \mu}  K_{ j \nu} K^{\rho}_{k}  + {\nabla}_{\rho} K_{ i \nu}  K_{ j \mu} K^{\rho}_{k}) \left( c_{l_1 l_2}^{~~~i} r^{j l_1} r^{k l_2} + c_{l_1 l_2}^{~~~k} r^{i l_1} r^{j l_2} + c_{l_1 l_2}^{~~~j} r^{k l_1} r^{i l_2}\right) = 0, 
\eea
where we have used symmetries of the curvature terms and Killing identities (see appendix for more details).  This again, gives the CYBE \eqref{CYBE}.

\paragraph{Third order:}
To work out equations at the third order, we recall \eqref{g-B-phi-expanded} and that $g_{\mu\nu}$ and $\phi$ have even powers of $\Theta$ while the NSNS two-form has odd powers and hence $X$ has all powers from zero to three. Therefore, only the $X$-terms in the dilaton and Einstein equations contribute to third order. One may show that these equations become an identity once we use the fact that $I$ is a Killing vector, namely, ${\cal L}_I \Phi={\cal L}_I G_{\mu\nu}={\cal L}_I\Theta=0$. 

The only non-trivial equation at third order is hence the NSNS two-form equation. Again, after lengthy but straightforward analysis, one finds that this equation upon using $I$ being a Killing yields the CYBE. 

\paragraph{Higher orders:} From  \eqref{g-B-phi-expanded} one can readily see the following structure: For \emph{even powers of $\Theta$} the NSNS two-form EOM is satisfied trivially if $I$ is a Killing vector, while the dilaton and Einstein equations are non-trivial. Conversely, for \textit{odd powers of $\Theta$}, the dilaton and Einstein equations are readily satisfied once we assume $I$ is Killing. Given our analysis above, we expect the dilaton and Einstein equations at even powers, and the B-field EOM at odd powers yield the CYBE. It is, of course, desirable to provide such an analysis and give an all-orders proof for our proposed ``YB solution generating technique'', but we leave this to future work. It is clear that unless one can work by induction, perturbative expansions are not a means to provide such a proof.

\section{Examples}
\label{sec:ex}
In this section, we provide examples of generalised  Yang-Baxter deformations in a bid to get the reader better acquainted with the solution generating technique outlined in \cite{Bakhmatov:2017joy}. We focus on two examples that {fall outside the usual examples studied via} the Yang-Baxter $\sigma$-model, before presenting a more familiar example with an RR sector. 

\subsection{Flat spacetime}
Let us consider flat spacetime in three dimensions 3D. One may imagine that this is trivial compared to deformations of AdS spacetimes, but it turns out that generalising  the Yang-Baxter $\sigma$-model to flat spacetime is complicated by the fact that the bilinear of the coset Poincar\'e group is degenerate \cite{Matsumoto:2015ypa}. As a result,  our analysis here, simple though it may be, is novel. 

Consider the metric,  
\be
\dd s^2 = - \dd t^2 + \dd x^2 + \dd y^2. 
\ee
Since we will initially study the CYBE, we identify the isometry group of the spacetime. Flat spacetime is a maximally symmetric space and for this reason it permits six Killing vectors in 3D. Let us label the Killing vectors as follows:
\bea
T_1 &=& \partial_{t}, \quad T_2 = \partial_{x}, \quad T_3 = \partial_{y}, \nn
T_4 &=& t \partial_x + x \partial_{t}, \quad T_5 =  t \partial_y + y \partial_{t}, \quad T_6 = x \partial_y - y \partial_{x}, 
\eea
and record the non-zero commutation relations: 
\bea
\left[ T_1, T_4 \right] &=& T_2, \quad \left[ T_1, T_5 \right] = T_3, \quad 
\left[ T_2, T_4 \right] = T_1, \quad \left[ T_2, T_6 \right] = T_3, \quad \left[ T_3, T_5 \right] = T_1, \nn 
\left[ T_3, T_6 \right] &=& -T_2, \quad 
\left[ T_4, T_5 \right] = T_6, \quad \left[ T_4, T_6 \right] = T_5, \quad \left[ T_5, T_6 \right] = - T_4. 
\eea 

Let us consider the candidate $r$-matrix 
\be
\label{r_lorentz}
r = \alpha T_4 \wedge T_5 + \beta T_5 \wedge T_6 + \gamma T_6 \wedge T_4, 
\ee
where $\alpha, \beta$ and $\gamma$ are constant coefficients. We have deliberately picked the Lorentz generators, as once combined with translations one can easily generate more involved $r$-matrices through inner automorphisms of the algebra. As we shall see, inner automorphisms correspond to coordinate transformations in the geometry. Identifying $r^{45} = \alpha, r^{56} = \beta$, etc, we can substitute them into the CYBE (\ref{CYBE}), to identify a single constraint on the coefficients:
\be
\label{CYBE_lorentz} 
\alpha^2  = \beta^2 + \gamma^2. 
\ee
Given our earlier analysis, it can be expected that the same condition arises from the EOMs of generalized supergravity. To see this, we first recast the $r$-matrix as $\Theta$, using $\Theta = r$, 
\be
\Theta = \alpha (t \partial_x + x \partial_{t}) \wedge (t \partial_y + y \partial_{t}) + \beta (t \partial_y + y \partial_{t}) \wedge ( x \partial_y - y \partial_{x}) + \gamma ( x \partial_y - y \partial_{x}) \wedge (t \partial_x + x \partial_{t}), 
\ee
where we we have replaced the generators $T_i$ by their Killing vector representation. Having done so, one can easily read off the components of $\Theta$,
\bea
\label{theta_lorentz}
\Theta^{tx} &=& -y \Delta, \quad \Theta^{ty} = x \Delta, \quad \Theta^{xy} = t \Delta,  
\eea
where we have defined $\Delta = \alpha t + \beta y - \gamma x$. One can determine the corresponding Killing vector from (\ref{I_theta}), 
\be
I = \alpha ( x \partial_y - y \partial_x ) - \beta (t \partial_x + x \partial_t  )  - \gamma ( y \partial_t  + t \partial_y ).  
\ee 

At this stage, one generates a deformed supergravity solution from (\ref{map}) and (\ref{dilaton}), 
\bea
g_{\mu \nu} \dd x^{\mu} \dd x^{\nu} &=& \frac{1}{[1 + \Delta^2(t^2 - x^2 - y^2)]} \biggl[ - \dd t^2 + \dd x^2 + \dd z^2 \nn 
&-& \Delta^2 \left[ (t \dd t - x \dd x)^2 + (t \dd t - y \dd y)^2 + (x \dd x + y \dd y)^2 \right]  \biggr], \nn
B &=& - \frac{\Delta}{[1 + \Delta^2(t^2 - x^2 - y^2)]} \left( y \dd t \wedge \dd x + t \dd x \wedge \dd y + x \dd y \wedge \dd t\right), \nn
\phi &=& - \frac{1}{2} \log \left[ 1 + \Delta^2 (t^2 - x^2 - y^2) \right]. 
\eea
It is interesting to look at the symmetries preserved by the deformation. Obviously, there is one Killing vector $I$ that is a linear combination of the Lorentz transformations generated by $T_4, T_5$ and $T_6$. Plugging the deformed geometry, along with $I$, into the EOMs of generalized supergravity, one quickly confirms that a deformation exists provided (\ref{CYBE_lorentz}) holds, in line with our expectations. It is worth noting that one can easily generate more complicated solutions by shifting $t, x$ and $y$ by constants, since translation symmetries are broken. 

The point of this example is to demonstrate the the EOMs are equivalent to the CYBE. However, if the focus is on inequivalent $r$-matrix solutions to the CYBE, we note that $r$-matrices related through inner automorphisms of the algebra are equivalent. Therefore, by applying an inner automorphism to (\ref{r_lorentz}), we can bring it to the simpler form, 
\be
r = \alpha ( T_4 \wedge T_5 + T_5 \wedge T_6).  
\ee
To see that the $r$-matrices are equivalent, note that one can generate an $r$-matrix satisfying (\ref{CYBE_lorentz}) through the inner automorphism $e^{\theta T_6} X e^{-\theta T_6}$, where $X \in \{ T_4, T_5, T_6 \}$ and $\beta = \alpha \cos \theta, \gamma = - \alpha \sin \theta$. The inner automorphism of the algebra corresponds to a rotation by angle $\theta$ in the $(x,y)$-plane.

\subsection{Bianchi III}
The previous example involved a deformation of flat spacetime. In a bid to consider spacetimes that are not Ricci-flat, let us consider the following Bianchi III spacetime,  
\bea
\dd s^2 &=& - a_1^2 a_2^2 a_3^2 e^{-4 \phi} \dd t^2 + a_1^2 \sigma_1^2 + a_2^2 \sigma_2^2 + a_3^2 \sigma_3^2, \nn
\Phi &=& \lambda t, 
\eea
where we have defined the functions 
\be
a_1 = a_3 = \frac{p_1}{\sinh (p_1 t)} e^{- \frac{1}{2} p_2 t + \lambda t}, \quad a_2 = e^{\frac{1}{2} p_2  t + \lambda t},  
\ee
and Maurer-Cartan one-forms: 
\be
\sigma_1 = \dd x, \quad \sigma_2 = \dd y, \quad \sigma_3 = e^{x} \dd z. 
\ee
A solution exists provided the constants satisfy the condition: 
\be
4 p_1^2 = p_2^2 + 4 \lambda^2. 
\ee
The Killing vectors are 
\be
T_1 = \partial_{x} - z \partial_{z}, \quad T_2 = \partial_y, \quad T_3 = \partial_z, 
\ee
and they satisfy the commutation relation: 
\be
[T_1, T_3 ] = T_3. 
\ee
Given that we only have three Killing vectors, the most general $r$-matrix one can consider is 
\be
r = \alpha T_1 \wedge T_2 + \beta T_2 \wedge T_3 + \gamma T_3 \wedge T_1. 
\ee
It is easy to check that this is a solution to the CYBE (\ref{CYBE}) provided $\alpha \gamma = 0$. Here, the $\beta$ term corresponds to a naive TsT transformation in the $(y,z)$-directions, both of which are Killing. Since the TsT deformation is of less interest, we will henceforth consider $\beta = 0$. 

It remains now to check that the EOMs agree with the CYBE and that valid solutions exist when either $\alpha \neq 0 $ or $\gamma \neq 0$. As before, we extract the components of $\Theta$, 
\be
\Theta^{x y} = \alpha, \quad \Theta^{yz} =  \alpha z,\quad \Theta^{z x} = \gamma.
\ee
and identify the corresponding Killing vector, 
\be
I  = - \gamma \partial_{z}
\ee
When $\gamma =0$, it can be checked that the deformed geometry 
\bea
g_{\mu \nu} \dd x^{\mu} \dd x^{\nu} &=& - a_1^2 a_2^2 a_3^2 e^{-4 \lambda t} \dd t^2  + \frac{1}{[1 + \alpha^2 a_2^2 (a_1^2 + z^2 e^{2 x} a_3^2)]} \biggl[ a_1^2 \dd x^2 \nn 
&+& a_2^2 \dd y^2 + a_3^2 e^{2 x} \dd z^2  + \alpha^2 e^{2 x} a_1^2 a_2^2 a_3^2 ( z \dd x + \dd z)^2 \biggr], \nn
B &=& - \frac{\alpha a_2^2}{[1 + \alpha^2 a_2^2 (a_1^2 + z^2 e^{2 x} a_3^2)]} ( a_1^2 \dd x \wedge \dd y + z e^{2 x} a_3^2 \dd y \wedge \dd z ), \nn
\phi &=& \lambda t - \frac{1}{2} \log [ 1 + \alpha^2 a_2^2 (a_1^2 + z^2 e^{2 x} a_3^2)],  
\eea
is a solution to usual supergravity. Since we have encountered a solution to usual supergravity, this deformation can be interpreted as a TsT transformation with respect to the shift symmetries generated by $T_1$ and $T_2$, respectively. It should be noted that both of these Killing vectors commute and the $r$-matrix is Abelian, so it is a TsT transformation \cite{Osten:2016dvf}. 

On the other hand, setting $\alpha = 0$, we encounter the geometry
\bea
g_{\mu \nu} \dd x^{\mu} \dd x^{\nu} &=& - a_1^2 a_2^2 a_3^2 e^{-4 \lambda t} \dd t^2 + a_2^2 \dd y^2 + \frac{1}{[ 1+ e^{2 x} \gamma^2 a_1^2 a_3^2]} \left[ a_1^2 \dd x^2 + a_3^2 e^{2 x} \dd z^2 \right], \nn
B &=& \frac{\gamma e^{2 x} a_1^2 a_3^2}{[ 1+ e^{2 x} \gamma^2 a_1^2 a_3^2]} \dd x \wedge \dd z, \nn
\phi &=& \lambda t - \frac{1}{2} \log [ 1+ e^{2 x} \gamma^2 a_1^2 a_3^2]. 
\eea
It is straightforward to check that the EOMs of generalized supergravity are satisfied. This deformation is of Jordanian type. 

\subsection{Lunin-Maldacena-Frolov} 
As promised we give one example of a geometry with an RR sector simply to illustrate the utility of the methods outlined in \cite{Bakhmatov:2017joy}. While it is easy to consider a new example, and we invite readers to do so, this risks distracting the reader from our main message. For this reason, we find it instructive to study an example familiar to all. The key take-home message is that one can now perform a complicated series of TsT transformations in the NS sector by simply inverting a matrix, while the transformation of the RR sector follows from a knowledge of the bivector $\Theta$ and the Page-forms \cite{Page:1984qv} of the original geometry, as discussed in \cite{Araujo:2017jap,Araujo:2017enj}. While we do not provide a proof of the transformation of the RR sector, it can be checked case by case that it works and it is expected to work since all information about the deformation has to be encoded in $\Theta$. Our prescription for the RR sector transformation, which we review below is an elegant way to generate new fluxes solely based on a knowledge of $\Theta$. 

We focus on a well-known geometry that can be generated through a series of TsT transformations, namely the Lunin-Maldacena-Frolov geometries \cite{Lunin:2005jy, Frolov:2005dj}. As the reader will observe, while the deformation is traditionally defined in terms of a series of T-duality transformations, using our prescription this is a single matrix inversion: there is no need to return to the Buscher T-duality rules. We begin by recalling the original undeformed geometry:
\bea
\dd s^2 &=& R^2 \left( \dd s^2 (AdS_5) + \sum_{i=1}^3 (\dd r_i^2 + r_i^2 \dd \phi_i^2 ) \right), \nn
F_5 &=& 4 R^4 \left[ \vol (AdS_5) + \vol (S^5) \right],  
\eea
where we have introduced the constrained coordinates $r_i$, 
\be
r_1 = \cos \alpha, \quad r_2 = \sin \alpha \cos \theta, \quad r_3 = \sin \alpha \sin \theta. 
\ee
To find the deformed background using our method, it is easiest to work with the constrained coordinates. In terms of these coordinates the matrix to be inverted to get $g, B$, may be written as 
\be
G^{-1} + \Theta   = \left( \begin{array}{cccccc} 
R^{-2} & 0 & 0 & 0 & 0 & 0 \\  
0 & R^{-2} r_1^{-2} & 0 & \gamma_3 & 0 & - \gamma_2 \\
0 & 0 & R^{-2} & 0 & 0 & 0 \\
0 & -\gamma_3 & 0 & R^{-2} r_2^{-2} & 0 & \gamma_1 \\
 0& 0 & 0 & 0 & R^{-2} & 0 \\
0 & \gamma_2  & 0 & -\gamma_1 & 0 & R^{-2} r_3^{-2}
\end{array} \right),  
\ee
where we have labeled the columns and rows $r_1, \phi_1, r_2, \phi_2$, etc. Inverting this matrix, while redefining $\hat{\gamma}_i = R^2 \gamma_i$, we get the following metric and NSNS two-form: 
\bea
\dd s^2 &=& R^2 \left[ \dd s^2 (AdS_5) + \sum_{i=1}^3 ( \dd r_i^2 + G r_i^2 \dd \phi_i^2) + G r_1^2 r_2^2 r_3^2 \left( \sum_{i=1}^3 \hat{\gamma}_i \dd \phi_i \right)^2 \right], \nn
B &=& - R^2 G \left( \hat{\gamma}_3 r_1^2 r_2^2 \dd \phi_1 \wedge \dd \phi_2 + \hat{\gamma}_1 r_2^2 r_3^2 \dd \phi_2 \wedge \dd \phi_3 + \hat{\gamma}_2 r_3^2 r_1^2 \dd \phi_3 \wedge \dd \phi_1\right) , 
\eea
where we have defined 
\bea
G^{-1} &=&  1 +  \hat{\gamma}_3^2 r_1^2 r_2^2 + \hat{\gamma}_1^2 r_2^2 r_3^2 + \hat{\gamma}_2^2 r_3^2 r_1^2. 
\eea
It is easy to check that this is, up to a sign in the NSNS two-form, the usual form of the solution. The dilaton is read off from the T-duality invariant $e^{-2 \phi} \sqrt{-g}$, leading to 
\be
e^{2 \phi} = G, 
\ee
and it can be checked that $I=0$, so we find a bona fide supergravity solution, as expected. 

Before illustrating how the RR sector transforms, let us review the logic. Since the Page charges are quantised, we do not expect them to change under the deformation since $\gamma_i$ are continuous deformation parameters. Therefore, the Page five-form should be invariant. We note that from the perspective of AdS/CFT, this invariance is very natural as the Page charges, which arise from integrating the Page forms over compact cycles, are related to the ranks of the gauge groups. Using the invariance of the Page form, we can get the lower dimension Page forms by simply contracting the bivector $\Theta$ and its products into the invariant Page forms. This procedure works for all the geometries we have considered, so we expect it to work in this setting too, and we will quickly confirm it does. 

To extract the RR sector, we define the Page forms in terms of the usual RR field strengths, 
\be
\label{eq}
Q_1 = \tilde{F}_1, \quad Q_3 = \tilde{F}_3- B \wedge \tilde{F}_1, \quad Q_5 = \tilde{F}_5 - B \wedge \tilde{F}_3,  
\ee
where we have added tildes to distinguish the deformed RR sector from the original RR sector. We have also flipped the sign of the $B$-field relative to \cite{Bakhmatov:2017joy} to make our conventions consistent with \cite{Lunin:2005jy, Frolov:2005dj}. As explained above, we now use the fact that $Q_5$ is invariant, which implies it is the same as the original five-form flux,
\be
Q_5 = F_5 = 4 R^4 \left( \vol (AdS_5) + \vol (S^5) \right), 
\ee 
since there was no NSNS two-form in the beginning. 

Our prescription \cite{Bakhmatov:2017joy} now demands that we contract in $\Theta$ and its products to get the lower-dimensional Page forms. It should be noted that all products of $\Theta$ vanish when contracted into forms, so we only need to contract $\Theta$ to find the Page three-form $Q_3$ with the Page one-form being trivially zero.  As a result, we have  
\be
\tilde{F}_{3 \, \rho_1 \rho_2 \rho_3} = Q_{3 \, \rho_1 \rho_2 \rho_3}  = \frac{1}{2!} \Theta^{\mu \nu} Q_{5 \,  \mu \nu \rho_1 \rho_2 \rho_3} . 
\ee
Following this procedure, we get 
\be
\tilde{F}_3 = 4 R^2 \sin^3 \alpha \cos \alpha \sin \theta \cos \theta \dd \alpha \wedge \dd \theta \wedge \sum_{i=1}^3 \hat{\gamma}_i \dd \phi_i. 
\ee
Now that we have $B$ and $\tilde{F}_3$, we can read off $\tilde{F}_5$ from (\ref{eq}). The result is 
\be
\tilde{F}_5 = 4 R^4 \left[ \vol (AdS_5) + G \vol(S^5) \right]. 
\ee
Up to signs, our expressions for the RR sector agree with \cite{Lunin:2005jy, Frolov:2005dj}. We emphasise again that there was no need to perform any T-duality transformation or use results in the literature detailing how the RR sector transforms \cite{Sakamoto:2018krs, Borsato:2016ose}. It is much quicker to get the RR sector using invariance of the Page forms and descent through $\Theta$ contractions, as was also checked earlier in \cite{Araujo:2017jap, Araujo:2017enj}.

\section{Discussion} 
In this work, we focused on the generalized supergravity EOMs and analysed what they imply on solutions obtained from deformations generated through the open-closed string map, and in this way, substantiated the claims of our earlier letter \cite{Bakhmatov:2017joy}. Assuming the bivector $\Theta$ to be a generic linear combination of anti-symmetric products of Killing vectors, we imposed the EOMs of generalized supergravity and studied the equations perturbatively in $\Theta$. Our analysis revealed:\begin{enumerate}
\item The consistency (integrability condition) of generalized supergravity EOMs implies the form of the $X$ field appearing in these equations in terms of $I$ and other fields. 
\item The $I$ vector appearing in the generalized supergravity must be a Killing vector of the deformed background and is also the divergence of $\Theta$. This generalises the earlier unimodularity condition of \cite{Borsato:2016ose}, which is recovered when $I =0$. It also proves the observation made through a long list of examples \cite{Araujo:2017jkb,Araujo:2017jap,Araujo:2017enj, Sakamoto:2018krs, Hong:2018tlp}.
\item Most importantly, the CYBE comes out as a result of the EOMs and not as an input.
\end{enumerate}
 We have hence provided strong evidence that the CYBE and our open-closed map can be used beyond coset or maximally symmetric spaces. We have then checked the  ``YB solution generating technique'' proposal in various examples.
Here we showed that the EOMs are automatically satisfied at the third order in $\Theta$ once the CYBE is imposed. However,  based on explicit examples, either presented earlier in \cite{Bakhmatov:2017joy}, or fleshed out in section \ref{sec:ex}, it should be clear that this statement is true for all orders. That being said, the proof of the YB solution generating technique is still outstanding. 

Since we were largely working perturbatively, but ultimately are interested in generating exact solutions to supergravity, we opted not to address the RR sector. Admittedly, if one is only working perturbatively, there is little motivation to do so. Our experience with many examples \cite{Araujo:2017jkb,Araujo:2017jap,Araujo:2017enj, Sakamoto:2018krs, Hong:2018tlp} indicates that the addition of the RR sector should largely be a technical issue and would just confirm the results we have enumerated above. The only new feature we expect to appear with the addition of the RR sector is the possibility of obtaining the modified CYBE; note that as we showed, the NS sector yields only the homogeneous CYBE. To obtain the modified CYBE within our framework, in \cite{Bakhmatov:2017joy} it was proposed to make an extra constant shift in the dilaton. However, given that the only known deformed geometries based on the modified CYBE are deformations of $AdS_p \times S^p$ geometries, we can use our approach to study deformations of Minkowski vacua supported by RR flux. We will report on this elsewhere. 

Our method, open-closed map plus solutions to CYBE, can be used as a very handy and simple solution generating proposal, as outlined in \cite{Bakhmatov:2017joy}. To demonstrate this, we reworked the Lunin-Maldacena-Frolov geometries to highlight the economy of the approach. Our method provides clearly a smarter way to perform TsT transformations, rather than going through the standard Buscher T-duality procedure.

As a final remark, we point out that in this work we focused on original backgrounds without any $B$-field; the $B$-fields that appear in the solutions are all generated through $\Theta$. There are, however, interesting geometries, such as $AdS_3\times S^3\times T^4$, which are supported by $H$-flux (briefly commented on in \cite{Araujo:2017jap, Sakamoto:2018krs}). For this example, the matrix $g+B$ is singular and cannot be inverted. In short, our method does not work. Nevertheless, one can consider the more general framework of $O(d,d)$ and $\beta$-transformations \cite{Rennecke:2014sca} (also \cite{Lust:2018jsx}), which include both non-Abelian T-duality and Yang-Baxter deformations as special cases \footnote{We thank Y. Sakatani and J. Sakamoto for correspondence on this issue.}.

\section*{Acknowledgements}

We thank T. Araujo, \"O. Kelekci, J. Sakamoto, Y. Sakatani and S. van Tongeren for discussion on related topics.  I. B. is partially supported by the Russian Government program for the competitive growth of Kazan Federal University. E. \'O C. thanks Cafe Dudart, Sangam-dong, Seoul for refuge. 
M.M. Sh-J. is partially supported by the grants from ICTP NT-04, INSF grant No 950124.  The work of HY is supported in part by National Natural Science Foundation of China, Project 11675244.

\appendix
\section{Consistency of the generalized supergravity field equations}

To check the consistency of equations (\ref{B_eom}-\ref{Einstein}), we first rewrite the Einstein equation (\ref{Einstein}) as follows
\be
\hat R_{\mu\nu}-\frac12g_{\mu\nu}\hat R=\frac14 (H_{\mu\alpha\beta}H_\nu^{\;\;\alpha\beta}-\frac12 H^2 g_{\mu\nu})-(\hat\nabla_\mu X_\nu+\hat\nabla_\nu X_\mu-g_{\mu\nu} \hat\nabla\cdot X). 
\ee
Taking the derivative of the above equation and noting that  $\hat\nabla^\mu(\hat R_{\mu\nu}-\frac12g_{\mu\nu}\hat R)=0$, we get
\be\frac14 H_\nu^{\;\;\alpha\beta} \hat\nabla^\mu H_{\mu\alpha\beta} +\frac14  H_{\mu\alpha\beta} \hat\nabla^\mu H_\nu^{\;\;\alpha\beta}-\frac18\hat\nabla_\nu H^2-\hat\nabla^2X_\nu-\hat\nabla^\mu\hat\nabla_\nu X_\mu+\hat\nabla_\nu \hat\nabla\cdot X=0. 
\ee
We also record the following,  
\be
\hat\nabla^\mu \hat\nabla_\nu X_\alpha=\hat\nabla_\nu\hat\nabla^\mu X_\alpha + X_\beta \hat R^{\beta\;\;\;\;\mu}_{\;\; \alpha\nu}\quad \Longrightarrow\quad \hat\nabla^\mu \hat\nabla_\nu X_\mu=\hat\nabla_\nu\hat\nabla\cdot X + X_\beta \hat R^{\beta}_{\;\;\nu}, 
\ee
\be\frac14 H_\nu^{\;\;\alpha\beta} \hat\nabla^\mu H_{\mu\alpha\beta} +\frac14  H_{\mu\alpha\beta} \hat\nabla^\mu H_\nu^{\;\;\alpha\beta}-\frac18\hat\nabla_\nu H^2-\hat\nabla^2X_\nu-X_\beta \hat R^{\beta}_{\;\;\nu}=0. 
\ee
One may also note that
\bea
 &&\hspace{-25mm}H_{\mu\alpha\beta} \hat\nabla^\mu H_\nu^{\;\;\alpha\beta}=H_{\mu\alpha\beta} \hat\nabla_\nu H^{\alpha\beta\mu}-H_{\mu\alpha\beta}\hat\nabla^\alpha H^{\beta\mu}_{\;\;\;\;\nu}+H_{\mu\alpha\beta}\hat\nabla^\beta H^{\mu\;\;\;\alpha}_{\;\;\;\nu},  \nonumber \\
&&=  H_{\mu\alpha\beta} \hat\nabla_\nu H^{\alpha\beta\mu}
-H_{\alpha\mu\beta}\hat\nabla^\mu H^{\beta\alpha}_{\;\;\;\;\nu}
+H_{\beta\alpha\mu}\hat\nabla^\mu H^{\beta\;\;\;\alpha}_{\;\;\;\nu}, 
\nonumber \\
&& = H_{\mu\alpha\beta} \hat\nabla_\nu H^{\alpha\beta\mu}
-H_{\mu\alpha\beta}\hat\nabla^\mu H_\nu^{\;\;\;\alpha\beta}
-H_{\mu\alpha\beta}\hat\nabla^\mu H_\nu^{\;\;\;\alpha\beta}, 
\eea
so we infer the relation 
\be
H_{\mu\alpha\beta} \hat\nabla^\mu H_\nu^{\;\;\alpha\beta}= \frac16 \hat\nabla_\nu H^2. 
\ee

Using the above identity and the $B$-field and Einstein equations (\ref{B_eom}, \ref{Einstein}),  we get
\bea
\frac12 H_\nu^{\;\;\alpha\beta}\left( X^\mu H_{\mu\alpha\beta}+\hat\nabla_\alpha X_\beta- \hat\nabla_\beta X_\alpha \right) &-&\frac{1}{12}\hat\nabla_\nu H^2 -\hat\nabla^2X_\nu \\
&-&X_\beta \left(\frac14 H^{\beta\mu\alpha}H_{\nu\mu\alpha}-\hat\nabla^\beta X_\nu-\hat\nabla_\nu X^\beta\right)=0,  \nonumber
\eea
and 
\be\label{A8}
\frac14 H_\nu^{\;\;\alpha\beta} X^\mu H_{\mu\alpha\beta}+\frac12H_\nu^{\;\;\alpha\beta} (\hat\nabla_\alpha X_\beta-\hat\nabla_\beta X_\alpha)-\frac{1}{12}\hat\nabla_\nu H^2  -\hat\nabla^2X_\nu+ X_\mu \hat\nabla^\mu X_\nu+ \frac12 \hat\nabla_\nu X^2=0. 
\ee
Using the dilaton equation (\ref{dilaton_EOM}) to replace $\nabla_\nu H^2$ and the Einstein equation to replace $H_\nu^{\;\;\alpha\beta} X^\mu H_{\mu\alpha\beta}$ in  equation (\ref{A8}) we get
\be
\hat R_{\mu\nu}X^\mu+2X_\mu\hat\nabla^\mu X_\nu+\frac12H_\nu^{\;\;\alpha\beta}( \hat\nabla_\alpha X_\beta-\hat\nabla_\beta X_\alpha)+\hat\nabla_\nu\hat\nabla_\mu X^\mu-\hat\nabla_\nu X^2              -\hat\nabla^2X_\nu=0, 
\ee
and
\be
 -\hat\nabla^2X_\nu+\hat\nabla_\mu\hat\nabla_\nu X^\mu +\frac12H_\nu^{\;\;\alpha\beta} (\hat\nabla_\alpha X_\beta  -\hat\nabla_\beta X_\alpha)+2X^\mu\hat\nabla_\mu X_\nu-\hat\nabla_\nu X^2 =0. 
\ee
This can be written as
\be\label{eqf}
\hat\nabla_\mu f^{\mu\nu} = \frac12H^{\nu\alpha\beta}f_{\alpha\beta} +2X_\mu f^{\mu\nu} ,\quad {\mathrm{where}} \quad f_{\mu\nu} = \hat\nabla_\mu X_\nu-\hat\nabla_\nu X_\mu. 
\ee
One can now decompose the one-form $X$ as an exact form, a Killing form and an extra part normal to the Killing form
\be
X_\mu=\partial_\mu \phi +I_\mu+\lambda_\mu, 
\ee
so we arrive at 
\be
f_{\mu\nu}=\partial_\mu I_\nu-\partial_\nu I_\mu -(\partial_\mu \lambda_\nu-\partial_\nu \lambda_\mu ). 
\ee
We note that, when $B=0$ we get $I=0$ and arrive back to the usual supergravity equations with $X_\mu=\partial_\mu \phi$. From (\ref{eqf}) we can argue that $f_{\mu\nu}$ should contain some term proportional to $H$. Therefore we write $\partial_\mu \lambda_\nu-\partial_\nu \lambda_\mu=z^\alpha H_{\alpha\mu\nu}$, where $z$ is an $H$-independent vector field. Replacing this ansatz in (\ref{eqf}) , we find
\bea
&& \hat\nabla^2  I^\nu-\hat\nabla_\mu \hat\nabla^\nu I^\mu-  H^{\alpha\mu\nu} \hat\nabla_\mu z_\alpha -   z_\alpha \hat\nabla_\mu H^{\alpha\mu\nu}   -\frac12H^{\nu\alpha\beta} \hat\nabla_\alpha I_\beta+\frac12H^{\nu\alpha\beta} \hat\nabla_\beta I_\alpha \nonumber \\ &&+  \frac12 z^\rho H^{\nu\alpha\beta}   H_{\rho\alpha\beta} -2X_\mu \hat\nabla^\mu I^\nu+2X_\mu \hat\nabla^\nu I^\mu+  2X_\mu z_\rho H^{\rho\mu\nu} =0. 
\eea
Using the equations of motion again, we get
\bea
&& \hat\nabla^2  I^\nu-\hat\nabla_\mu \hat\nabla^\nu I^\mu-  H^{\alpha\mu\nu} \hat\nabla_\mu z_\alpha  
- 2z_\alpha X_\rho H^{\rho\nu\alpha} - 2z_\alpha \hat\nabla^\nu X^\alpha +  2z_\alpha \hat\nabla^\alpha X^\nu
   -\frac12H^{\nu\alpha\beta} \hat\nabla_\alpha I_\beta
     \\ && 
     +\frac12H^{\nu\alpha\beta} \hat\nabla_\beta I_\alpha 
   +  2 z^\alpha R^\nu_{\;\rho}+2 z_\alpha \hat\nabla^\nu X^\alpha +2 z_\alpha \hat\nabla^\alpha X^\nu -2X_\mu \hat\nabla^\mu I^\nu+2X_\mu \hat\nabla^\nu I^\mu+  2 z_\alpha X_\rho H^{\rho\nu\alpha} =0  \nonumber, 
   \eea
and after some further simplifications we get 
\bea
2 (z^\alpha - I^\alpha) R^\nu_{\;\;\alpha} -H^{\alpha\mu\nu} \hat\nabla_\mu (z_\alpha-I_\alpha)  
  +  4z_\alpha \hat\nabla^\alpha X^\nu 
    +4X_\alpha \hat\nabla^\nu I^\alpha =0 . 
\eea

Next, noting that 
\be
{\mathcal L}_I X^{\nu} = I^\alpha\hat\nabla_\alpha X^{\nu} +  X_{\alpha} \hat\nabla^\nu I^\alpha =0, 
\ee
we arrive at 
\bea
2 (z^\alpha - I^\alpha) R^\nu_{\;\;\alpha} -H^{\alpha\mu\nu} \hat\nabla_\mu (z_\alpha-I_\alpha)  
  +  4(z_\alpha-I_\alpha) \hat\nabla^\alpha X^\nu 
   =0 
\eea

This is an identity when $z=I$. This along with the fact that when $B=0$, the Killing vector $I$ vanishes, implies $\lambda_\mu=-B_{\mu\nu} I^\nu$. It should be noted that in writing this solution we have absorbed a total derivative in $\phi$.

\section{Details of the perturbative analysis}
In this section, we provide some details of the results quoted in the text. 

\paragraph{Some useful Killing identities.} In our perturbative analysis we have heavily used Killing vectors and their properties. So we start with some useful identities. Given a set of Killing vectors $K_i^\mu$, 
\be
\nabla_{\mu}K_{i \, \nu}+\nabla_{\nu}K_{i \, \mu}=0,
\ee
there is a well-known identity, 
\be
\label{R-K}
\nabla_{\alpha} \nabla_{\beta} K^{\gamma} = R^{\gamma}_{~\beta \alpha \lambda} K^{\lambda},\quad \Rightarrow \quad \nabla^2 K_\mu=-R_{\mu\nu} K^{\nu}\;.
\ee
Assuming the bi-Killing structure for $\Theta^{\mu \nu}$ \eqref{theta}, one can readily show:
\be
H_{\mu\nu\alpha}=\nabla_{\alpha}\Theta_{\mu\nu}+\nabla_{\nu}\Theta_{\alpha\mu}+\nabla_{\mu}\Theta_{\nu\alpha}=2r_{ij}\left[\nabla_{\alpha}K^i_\mu K^j_\nu+\nabla_{\mu}K^i_\nu K^j_\alpha+\nabla_{\nu}K^i_\alpha K^j_\mu\right], 
\ee
where the spacetime indices are lowered and raised by the metric $G_{\mu\nu}$.

\paragraph{Perturbative expansion.} 
Expanding the metric $g_{\mu \nu}$, $B$-field and dilaton for small $\Theta$, we get
\bea
&& g_{\mu\nu}=G_{\mu\nu}+\Theta_{\mu\rho}\Theta^\rho_{\;\;\nu}+{\mathcal O}(\Theta^4),\\
&& B_{\mu\nu}=-\Theta_{\mu\nu}-\Theta_{\mu\alpha}\Theta^{\alpha\beta} \Theta_{\beta\nu} +{\mathcal O}(\Theta^5),\nonumber \\ && \phi=\Phi+\frac14 \Theta_{\mu\nu}\Theta^{\mu\nu}+{\mathcal O}(\Theta^4)\nonumber. 
\eea
Plugging the above expressions directly into the EOMs, we can also expand them for small $\Theta$. We now detail the information extracted at each order from the EOMs. 
\subsection{Zeroth Order}
At zeroth order in $\Theta$, equations (\ref{B_eom}-\ref{Einstein}) become
\bea
\nabla_\mu X_\nu-\nabla_\nu X_\mu=0 &\Longrightarrow&  \nabla_\mu\nabla_\nu\Phi- \nabla_\nu\nabla_\mu\Phi=0 \; \; \;\; 
\\
R_{\mu\nu} +\nabla_\mu X_\nu +\nabla_\mu X_\nu=0 & \Longrightarrow & R_{\mu\nu}+2\nabla_\mu\nabla_\nu\Phi=0  \; \; \;\;\;\;\;\; 
\\
\nabla \cdot X -2 X^2=0 & \Longrightarrow & \nabla^2\Phi -2\nabla_\mu\Phi \nabla^\mu\Phi=0  \; \; \;\; 
\eea
where we remark that the first equation is trivial, whereas the second and third are simply the EOMs satisfied by the original undeformed solution, in line with expectations. 

\subsection{First Order}
At first order the linear terms in $\Theta$ give the following contribution to the EOMs, 
\bea
\label{H1st}
\frac12\nabla^\rho H_{\rho\mu\nu}-(\nabla^\rho\Phi) H_{\rho\mu\nu}-\nabla_\mu I_\nu+\nabla_\nu I_\mu &=&0,  \\
\label{E1st}
\nabla_\mu I_\nu +\nabla_\nu I_\mu &=&0, \\
\label{S1st}
\nabla_\mu I^\mu -4 \nabla_\mu \Phi I^\mu=0 \quad\Longrightarrow\quad I\cdot \nabla \Phi &=&0. 
\eea
Now we assume that $\Theta$ is bi-Killing (\ref{bi-Killing}) and use the identities (\ref{R-K}) 
\be
\nabla\cdot K=0,\qquad K\cdot \nabla \Psi=0, 
\ee
where the latter is valid for any field $\Psi$. This allows us to write $H= \dd B$ in terms of the components of $\Theta$ as
\bea
H_{\alpha\mu\nu} &=&r_{ij}\left(K_\nu^j\nabla_\alpha K_\mu^i+ K_\mu^i\nabla_\alpha K_\nu^j +K_\nu^i  \nabla_\mu K_\alpha^j +K_\alpha^j \nabla_\mu K_\nu^i   +K_\alpha^i \nabla_\nu K_\mu^j  +  K_\mu^j   \nabla_\nu K_\alpha^i    \right),  \nn
&=&2 r_{ij}\left(K_\nu^i\nabla_\mu K_\alpha^j+ K_\alpha^i \nabla_\nu K_\mu^j+ K_\mu^i\nabla_\alpha K_\nu^j  \right).  
\eea
Now each term in equation (\ref{H1st}) can be expanded as follows: the first term is 
\bea
 &&\hspace{-19mm}\frac{1}{2}\nabla^\alpha H_{\alpha\mu\nu}=
  r_{ij} \left(2 \nabla^\alpha K_\nu^j\nabla_\alpha K_\mu^i -2K_\nu^j  \nabla_{\mu} K_\gamma^i \nabla^{\gamma}\Phi -  2 K_\mu^i \nabla_{\nu} K_\gamma^j \nabla^{\gamma}\Phi  + R_{\nu\mu}^{\;\;\;\; \alpha\gamma} K_\alpha^j  K_\gamma^i  \right); 
\eea
the second term  is
\bea
&& \nabla^\gamma \Phi H_{\gamma\mu\nu}=
-2r_{ij}( K^j_\nu  \nabla_\mu K^i_\gamma \nabla^\gamma\Phi + K^i_\mu \nabla_\nu  K^j_\gamma \nabla^\gamma\Phi ), 
\eea
so that equation (\ref{H1st}) can be further simplified, 
\bea
&& r_{ij} \left(2 \nabla^\alpha K_\nu^j\nabla_\alpha K_\mu^i  +R_{\nu\mu}^{\;\;\;\; \alpha\gamma} K_\alpha^j  K_\gamma^i  \right) -\nabla_\mu I_\nu+\nabla_\nu I_\mu =0 
 \nonumber  \\
 &&
 \nabla_\mu\bigg( \nabla^\alpha \left(r_{ij} K_\alpha^i  K_\nu^j  \right) - I_\nu\bigg) =0  
 \nonumber  \\
 &&
 \nabla_\mu\left( \nabla^\alpha \Theta_{\alpha\nu}  - I_\nu\right) =0  \quad  \Rightarrow \quad I_\nu=\nabla^\alpha \Theta_{\alpha\nu} + const.
  \eea

Equation (\ref{E1st}) implies that $I$ is a Killing vector of the original metric $G_{\mu \nu}$ and therefore (\ref{S1st}) is automatically satisfied for the scalar field $\Phi$.

\subsection{Second Order}

Before trying to expand and solve the second order equations, it would be useful to simplify the EOMs using what we have found from the zeroth and the first order equations. Using the fact that $I^\mu$ is a Killing vector we find 
\be\label{LDphi}
{\mathcal L}_I\phi=I^\mu\partial_\mu \Phi =0, 
\ee
\be 
{\mathcal L}_Ig=\hat\nabla_\mu I_\nu+\hat\nabla_\nu I_\mu =0, 
\ee
\be 
{\mathcal L}_IB=I^\alpha\hat\nabla_\alpha B_{\mu\nu} + B_{\alpha\nu}\hat\nabla_\mu I^\alpha +B_{\mu\alpha}\hat\nabla_\nu I^\alpha =0. 
\ee
Using these expressions, we get
\bea
&&\hspace{-11mm}-\hat\nabla_\mu X_\nu+\hat\nabla_\nu X_\mu=-g_{\alpha\nu}\hat\nabla_\mu I^\alpha+g_{\mu\alpha}\hat\nabla_\nu I^\alpha 
+ (\hat\nabla_\mu B_{\nu\alpha}-\hat\nabla_\nu B_{\mu\alpha})I^\alpha +(B_{\alpha\nu}\hat\nabla_\mu I^\alpha-B_{\alpha\mu}\hat\nabla_\nu I^\alpha), \nonumber \\
&&\hspace{20mm}=-\hat\nabla_\mu I_\nu+\hat\nabla_\nu I_\mu 
+ (\hat\nabla_\mu B_{\nu\alpha}+\hat\nabla_\nu B_{\alpha\mu}+\hat\nabla_\alpha B_{\mu\nu})I^\alpha, \nonumber \\
&&\hspace{20mm}=-\hat\nabla_\mu I_\nu+\hat\nabla_\nu I_\mu + I^\alpha H_{\alpha\mu\nu},  
 \nonumber \\&&\hspace{20mm}=-2\partial_\mu I_\nu 
+ I^\alpha H_{\alpha\mu\nu}, 
\eea

and also 

\bea
-X^\rho H_{\rho\mu\nu}=-\partial^\rho\phi H_{\rho\mu\nu}-I^\rho H_{\rho\mu\nu} +B^\rho_{\;\;\;\alpha}I^{\alpha} H_{\rho\mu\nu}. 
\eea

Therefore, the $B$-field equation (\ref{B_eom}) becomes, 
\be
\label{Beqnew}\frac12\hat\nabla^\rho H_{\rho\mu\nu}-\partial^\rho\phi H_{\rho\mu\nu} +B^\rho_{\;\;\;\alpha}I^{\alpha} H_{\rho\mu\nu}  -2\partial_\mu I_\nu=0. 
\ee

 To expand covariant derivatives we note that $\hat\Gamma^a_{bc}$ can be expand in $\Theta$ and gives
  \be
  \label{CDC}
  \hat\Gamma^\alpha_{\beta \rho}=\Gamma^\alpha_{\beta \rho}+C^\alpha_{\beta \rho},\quad {\mathrm{where}}\quad C^\alpha_{\beta \rho}=\frac12G^{\alpha \sigma}(\nabla_\beta \Theta^2_{\rho \sigma} +\nabla_\rho\Theta^2_{\sigma\beta } -\nabla_\sigma\Theta^2_{\beta \rho}  ) +\mathcal{O}(\Theta^4), 
  \ee
and
\be\label{Ddef}
\hat{R}_{\mu\nu} =R_{\mu\nu}+D_{\mu\nu}=R_{\mu\nu}+\nabla_\alpha C^\alpha_{\mu\nu}-\nabla_\mu C^\alpha_{\alpha\nu}. 
\ee

It is clear from (\ref{Beqnew}) that there is no second order term in the $B$-field equation. 

\paragraph{The Einstein equations} 
The quadratic term in the Einstein equation can be organised as follows, 
\be
D_{\mu\nu}-\frac14H_{\mu\alpha\beta}H_{\nu}^{\;\;\alpha\beta}+\nabla_\mu(\Theta_{\nu\rho}I^\rho)+\nabla_\nu(\Theta_{\mu\rho}I^\rho)   +2 C^\alpha_{\;\;\mu\nu} \partial_\alpha\Phi=0, 
\ee
where $D$ denotes the second order terms in the Ricci tensor expansion, defined in (\ref{Ddef}). The first term can be expand as follows
\bea
&&D_{\mu\nu}=\frac12 \nabla^\alpha\nabla_\mu \Theta^2_{\alpha\nu}-\frac14\nabla^2\Theta_{\mu\nu}^2-\frac14\nabla_\mu\nabla_\nu \Theta^2 + (\mu\leftrightarrow\nu), \\
&&
\qquad= r_{ij} r_{mn} \bigg(2 K^{j\beta}  K_\beta^m K_\nu^n \nabla^\alpha\nabla_\mu K^i_\alpha 
                                     -K^{j\beta} K_\beta^m K_\nu^n \nabla^2 K^i_\mu  
   -K^{j}_{\beta} K^{m\alpha} K^{n\beta}\nabla_\mu\nabla_\nu  K^i_\alpha 
 \bigg) + (\mu\leftrightarrow\nu). \nonumber
\eea

The second term gives
\bea
&& -\frac14H_{\mu\alpha\beta}H_{\nu}^{\;\;\alpha\beta} = -\frac14(\nabla_\mu\Theta_{\alpha\beta}+\nabla_\alpha\Theta_{\beta\mu}+\nabla_\beta\Theta_{\mu\alpha})(\nabla_\nu\Theta^{\alpha\beta}+\nabla^\alpha\Theta^{\beta}_{\;\;\nu}+\nabla^\beta\Theta_\nu^{\;\;\alpha}), \nonumber \\
&& = -{r_{ij} r^{mn}}\bigg(
 K^j_\beta K^{n\beta} \nabla_\mu K_\alpha^i  \nabla_\nu K_{m\alpha} 
+K^j_\mu K^{n\beta} \nabla_\alpha K_\beta^i  \nabla_\nu K_{m\alpha} +K^j_\beta K^n_\nu \nabla_\mu K_\alpha^i  \nabla_\alpha K^{m\beta} 
\nonumber \\&&\qquad \qquad \quad - K^j_\beta K^{n\alpha} \nabla_\mu K_\alpha^i \nabla_\nu K_{m\beta}     + \frac12 K^j_\mu K^n_\nu \nabla_\alpha K_\beta^i \nabla_\alpha K^{m\beta} 
 \bigg) + (\mu\leftrightarrow \nu). 
\eea

The third and the fourth terms together can be simplified as follows
\bea
&& \nabla_\mu (\Theta_{\nu\rho} \nabla_\beta \Theta^{\beta\rho}) + \nabla_\nu (\Theta_{\mu\rho} \nabla_\beta \Theta^{\beta\rho}) = \nabla_\mu \Theta_{\nu\rho} \nabla_\beta \Theta^{\beta\rho} + \Theta_{\nu\rho}  \nabla_\mu \nabla_\beta \Theta^{\beta\rho} + (\mu\leftrightarrow \nu), \\
&&={r_{ij} r^{mn}}\bigg( K^i_\nu K_m^\beta \nabla_\mu  K^j_\rho  \nabla_\beta K_n^\rho       
+ K_\nu^m K_\beta^n \nabla_\mu K^\rho_j \nabla_\beta K^\rho_i  
+ K_\nu^m K^{n\rho}  K^\beta_i K^\alpha_j R_{\rho\beta\mu\alpha}
 \bigg) + (\mu\leftrightarrow \nu),  \nonumber
\eea
and finally the last part of the Einstein equations give
\bea
&& 2C^\alpha_{\mu\nu} \partial_\alpha\Phi = 2 {r_{ij} r^{mn}}\bigg( K^i_\nu K^j_\eta  K^{m\eta} \nabla_\mu K^{n\alpha}  + K^i_\mu K^j_\eta K^{m\eta} \nabla_\nu K^{n\alpha}     +K^{i\eta} K^{j}_{\;\;\nu} K^n_\eta    \nabla_\mu K^{m\alpha} \nonumber \\ && \hspace{35mm}
+ 
K^{i\eta} K^{j}_{\;\;\nu}  K^m_\mu\nabla_\eta K^n_\alpha  +   K^i_\mu K^{j\eta} K^{n}_{\;\;\nu}\nabla_\eta K^{m\alpha} +   
K^i_\mu K^j_\eta K^{m\eta} \nabla_\nu K^{n\alpha} \bigg)\partial_\alpha\Phi,  
\nonumber 
\\
&& \hspace{15mm}= {r_{ij} r^{mn}} \bigg (2 K^i_\nu K^j_\eta  K^{m\eta}  K^{n\alpha} R_{\mu\alpha}  + 
K^{i\rho} K^{j}_{\;\;\nu}  K^m_\mu K^n_\alpha   R_{\rho\alpha}    \bigg) + (\mu\leftrightarrow \nu). 
\eea

Adding all these terms together we get

\bea
&&\hspace{-17mm}  {r_{ij} r^{mn}} 
\bigg [ 2 K^{j\beta}  K_\beta^m K_\nu^n \nabla^\alpha\nabla_\mu K^i_\alpha 
 -K^{j\beta} K_\beta^m K_\nu^n \nabla^2 K^i_\mu  
   -K^{j}_{\beta} K^{m\alpha} K^{n\beta}\nabla_\mu\nabla_\nu  K^i_\alpha
 -K^j_\beta K^{n\beta} \nabla_\mu K_\alpha^i  \nabla_\nu K_{m\alpha} \nonumber \\ &&
-K^j_\mu K^{n\beta} \nabla_\alpha K_\beta^i  \nabla_\nu K_{m\alpha} 
-K^j_\beta K^n_\nu \nabla_\mu K_\alpha^i  \nabla_\alpha K^{m\beta} 
 + K^j_\beta K^{n\alpha} \nabla_\mu K_\alpha^i \nabla_\nu K_{m\beta}  
 - \frac12 K^j_\mu K^n_\nu \nabla_\alpha K_\beta^i \nabla_\alpha K^{m\beta}   \nonumber \\ &&
+K^i_\nu K_m^\beta \nabla_\mu  K^j_\rho  \nabla_\beta K_n^\rho       
+ K_\nu^m K_\beta^n \nabla_\mu K^\rho_j \nabla_\beta K^\rho_i  
+ K_\nu^m K^{n\rho}  K^\beta_i K^\alpha_j R_{\rho\beta\mu\alpha}
+ 2 K^i_\nu K^j_\eta  K^{m\eta}  K^{n\alpha} R_{\mu\alpha}   \nonumber \\ &&+ 
K^{i\rho} K^{j}_{\;\;\nu}  K^m_\mu K^n_\alpha   R_{\rho\alpha}     
\bigg]+ (\mu\leftrightarrow \nu)=0. 
\eea
After simplification, this can be written as
  \bea
&&r_{ij}r_{mn}  \bigg[ K_{ i \mu} K^{\alpha}_{m} {\nabla}_{\alpha}K_\beta^j\nabla^\beta K_\nu^n   - K_{ i \mu} K^{\alpha}_{m} {\nabla}_{\alpha} K_\beta^n\nabla^\beta K_\nu^j  
 + K^{ m \mu} K^j_\alpha\nabla_\alpha K_\beta^n {\nabla}_{\beta} K^{ i \nu}  \nonumber\\
   &&  \qquad\quad -K^{ m \mu} K_\alpha^n\nabla_\alpha K^j_\beta  {\nabla}_{\beta} K^{ i \nu}  
    + K^j_\alpha K^{\beta}_{i} {\nabla}_{\beta} K^{ m \nu}\nabla_\alpha K_\mu^n 
  -K_\alpha^n K^{\beta}_{i} {\nabla}_{\beta} K^{ m \nu} \nabla_\alpha K^j_\mu  \nonumber\\
    && \qquad\quad - K_{ i \mu} K^{\alpha}_{m}K_\beta^n         R_{\mu\alpha\beta\nu}K_\nu^j   
   + K_{ i \mu} K^{\alpha}_{m}K_\beta^j  R_{\mu\beta\alpha\nu} K_\nu^n \bigg] + (\mu\leftrightarrow \nu)=0, 
 \eea
which upon further permutation of the indices, takes the neat form,  
\bea
(K^{j} _{\nu} K^{l\rho}  \nabla_\rho K^{ i}_{ \mu}   + K^{ i}_{ \nu}  K^{ j}_{\mu} K^{l\rho} \nabla_\rho) \left(  r_{j p} r_{l q} \;c^{pq}_{~~~i}+ r_{i p} r_{j q}  \;c^{pq}_{~~~l} +r_{lp} r_{iq}  \; c^{pq}_{~~~j}  \right)=0, 
\eea
which is identically satisfied one the CYBE holds.

\paragraph{Dilaton equation} Having discussed the more involved Einstein equation, we focus on the dilaton equation at second order. It takes the form, 

\be
\frac{1}{12}H^2+\nabla^\mu(\Theta_{\mu\nu} I^\nu)+C^\mu_{\;\;\;\mu\rho}\partial^\rho\Phi -4\partial^\mu\Phi \;\Theta_{\mu\nu}I^\nu-2I_\mu I^\mu=0. 
\ee

The second and the last term can be together simplified as
\bea
&&\nabla^\mu(\Theta_{\mu\nu} I^\nu)-2I_\mu I^\mu=\nabla^\mu \Theta_{\mu\nu}\nabla_\alpha\Theta^{\alpha\nu}+\Theta_{\mu\nu}\nabla^\mu\nabla_\alpha\Theta^{\alpha\nu}-2\nabla^\mu\Theta_{\mu\alpha}\nabla_\nu\Theta^{\nu\alpha},  \\ 
&&=
 r_{ij}r_{mn} \bigg(K_\mu^i K^{m\nu}\nabla^\mu K_\alpha^j  \nabla_\alpha K_\nu^n + K_\mu^i K_\nu^j \nabla^\mu K^{m\alpha}\nabla_\alpha K^{n\nu} +     K_\mu^i K_\nu^j  K^{m\alpha} \nabla^\mu \nabla_\alpha K^{n\nu}  \bigg). \nonumber 
\eea

Using equation (\ref{S1st}) and the fact that $\Theta$ is a bi-Killing, it is easily confirmed that $\Theta^{\mu\nu}\partial_\mu \Phi=0 $. The remaining term with $\Phi$ can be simplified as

 \be
 C^\mu_{\;\;\;\mu\rho}\partial^\rho\Phi=\frac12\left[\nabla_\mu (\Theta_{\rho}^{\;\; \mu})^2 +\nabla_\rho(\Theta^\mu_{\;\;\mu })^2 -\nabla^\mu\Theta^2_{\mu \rho}  \right]\partial^\rho\Phi =\frac12 \nabla_\rho(\Theta^\mu_{\;\;\mu })^2\partial^\rho\Phi, 
 \ee
where  $\Theta^2$ is defined as 
\be
\Theta^2_{\mu\nu}=\Theta_\mu^{~\alpha} \Theta_{\alpha\nu}. 
\ee
As a result, we get 
\bea
\frac12 \nabla_\rho(\Theta^\mu_{\;\;\mu })^2\partial^\rho\Phi&=&-\partial^\rho\Phi  \; (\nabla_\rho \Theta_{\mu\nu}) \Theta^{\mu\nu}, \\ 
&=&   -r_{ij}r_{mn}   K^{i}_{\nu}  K^{m\nu}  K^n_\alpha  \nabla^2 K^{j\alpha}.  \nonumber \nonumber
\eea
In the above we have used the B-field EOM to replace $\partial^\rho\Phi$ in terms of $\Theta$ and $K$.
The first term in the dilaton equation can hence be simplified as
\bea
&&\hspace{-12mm}\frac{1}{12}H^2 = \frac{1}{12}(\nabla_\mu\Theta_{\nu\alpha}+\nabla_\nu\Theta_{\alpha\mu}+\nabla_\alpha\Theta_{\mu\nu})(\nabla^\mu\Theta^{\nu\alpha}+\nabla^\nu\Theta^{\alpha\mu}+\nabla^\alpha\Theta^{\mu\nu}), \nonumber\\
&&=
 {r_{ij} r_{mn}}
\bigg( K^i_\nu  K^{m\nu}\nabla_\mu K^j_\alpha \nabla^\mu K^{n\alpha} 
+  2 K^n_\alpha K_i^\nu \nabla_\mu K_j^\alpha \nabla^\mu K_\nu^m
\bigg).  
   \eea

Putting everything together we get
\bea
&&{r_{ij} r_{mn}}
 \bigg(K_\mu^i K^{m\nu}\nabla^\mu K_\alpha^j  \nabla^\alpha K_\nu^n + K_\mu^i K_\nu^j \nabla^\mu K^{m\alpha}\nabla_\alpha K^{n\nu} +     K_\mu^i K_\nu^j  K^{m\alpha} \nabla^\mu \nabla_\alpha K^{n\nu}      \nonumber \\ && -K^{i}_{\nu}  K^{m\nu}  K^n_\alpha  \nabla^2 K^{j\alpha} +K^i_\nu  K^{m\nu}\nabla_\mu K^j_\alpha \nabla^\mu K^{n\alpha} 
+  2 K^n_\alpha K_i^\nu \nabla_\mu K_j^\alpha \nabla^\mu K_\nu^m
 \bigg) =0. 
 \eea
Further simplifying the above expression we get
 \bea
&& \bigg( K_{\mu}^i K^{m\nu}  \nabla^{\mu}  K_\alpha^j  \nabla^\alpha K_\nu^n
+  K_{\mu}^n K_{\nu}^j \nabla^{\mu}  K_\alpha^i \nabla^\nu K^{m\alpha}
+  K_{\mu}^m K_\nu^j \nabla^{\mu}    K_{\alpha}^{ i}     \nabla^\nu K^{n\alpha}   
-   2K_{\nu}^i K_\mu^j \nabla^\mu K_\alpha^n  \nabla^{\nu} K^{m\alpha} 
\nonumber\\
&& -   K_{\mu}^j K_\nu^n  \nabla^{\mu} K_{\alpha}^{ m} \nabla^\nu K^{i\alpha}    
   + K^i_\nu   K^j_\alpha     K^m_\mu  \nabla^\mu \nabla^\alpha    K^{n\nu}        
+  K^i_\mu K^j_\nu   K^m_\alpha       \nabla^\mu \nabla^\alpha     K^{n\nu}  
+  K^i_\alpha    K^j_\mu  K^m_\nu      \nabla^\mu \nabla^\alpha    K^{n\nu}  \nonumber
\\
&&
 - K_{\mu}^i K_{\nu}^m      K^n_\alpha  \nabla^{\mu}  \nabla^\alpha K^{j\nu}    
 +K_{\mu}^i K_{\nu}^m  K^j_\alpha   \nabla^{\mu}\nabla^\alpha K^{n\nu} \bigg) r_{ij}r_{mn}=0. 
 \eea
 
Now using  (\ref{R-K}) and the Bianchi identity for the Riemann tensor, we can factorise the above expression as 
\bea
&& K^{i\alpha} K^{l\beta} \nabla_{\alpha} K_{\beta}^m \left(r_{i p} r_{l q}  c^{p q}_{~~~ m} +  r_{m p} r_{i q} c^{p q}_{~~~ l}+  r_{l p} r_{m q}c^{p q}_{~~~ i} \right) \\
&&+ r_{ij}r_{mn}K^{i\beta} K^{j\mu} K^{m\alpha} K^{n\nu}\left( R_{\beta \mu \alpha \nu}  + R_{ \mu \alpha\beta \nu}  +R_{\alpha\beta\mu \nu}   \right) = 0. 
\eea
This expression vanishes identically for any curved background once the CYBE is satisfied.

  \subsection{Third order equations}
  To test that nothing funny happens at the higher order, we study the EOMs to third order in $\Theta$. We will see in this order that the dilaton and Einstein equation are somewhat trivial, while the $B$-field EOM encapsulates information of the CYBE. 
  
  \paragraph{Dilaton equations at third order}

In order to expand the EOMs, it is useful to note that
\be
\label{xlow}
X_\mu=\partial_\mu\phi + g_{\mu\nu}I^\nu-B_{\mu\nu}I^\nu = \partial_\mu\phi +\nabla^\alpha \Theta_{\alpha\mu} +\Theta_{\mu\nu}\nabla_\alpha\Theta^{\alpha\nu} +\Theta^2_{\mu\nu} \nabla_\alpha \Theta^{\alpha\nu}+ {\mathcal O}(\Theta^4). 
\ee
Therefore,  we have
\be
\label{xhigh}
X^\mu=g^{\mu\nu}X_\nu=\nabla^\mu\phi+\nabla_\alpha\Theta^{\alpha\mu}+\Theta^{\mu}_{~\beta}\nabla_\alpha\Theta^{\alpha\beta}+{\mathcal O}(\Theta^4),
\ee
where we have used (\ref{LDphi}). By expanding $H$ up to cubic order in $\Theta$,  we find
  \bea
  \label{H1H3}
 && H_{\rho\mu\nu}= H_{\rho\mu\nu}^{(1)}+H_{\rho\mu\nu}^{(3)}+\cdots \\ 
 &&\qquad\; =(\nabla_\rho\Theta_{\mu\nu} + \nabla_\mu\Theta_{\nu\rho}  + \nabla_\nu\Theta_{\rho\mu})+(\nabla_\rho\Theta_{\mu\nu}^3 + \nabla_\mu\Theta_{\nu\rho}^3  + \nabla_\nu\Theta_{\rho\mu}^3)+\mathcal{O}(\Theta^5), \nonumber
  \eea
 where we define
 \be
 \Theta^3_{\mu\nu}=\Theta_{\mu\alpha}\Theta^{\alpha\beta}\Theta_{\beta\nu}. 
 \ee
We note that the first term in the dilaton equation (\ref{dilaton_EOM}) does not have any term cubic in $\Theta$. Therefore, the dilaton equation reduces to, 
 \be
 C^\mu_{\mu\alpha}\nabla_\beta\Theta^{\beta\alpha}-2\Theta_{\mu\nu}\nabla_\beta\Theta^{\beta\nu}\nabla_\alpha\Theta^{\alpha\mu}+2\nabla_\alpha\Theta^{\alpha\beta}\nabla^\nu\Theta_{\mu\nu}\Theta^\mu_{\;\beta}=0. 
 \ee
 Using  (\ref{CDC}) and noting that $I$ is a Killing vector, we find that the above equation is satisfied identically.

 \paragraph{Einstein equation at third order}
At cubic order, the Einstein equation turns out to be the following:
\be
\nabla_\mu(\Theta_{\nu\beta}^2\nabla_\alpha\Theta^{\alpha\beta})-C^\alpha_{\mu\nu}\nabla^\beta\Theta_{\beta\alpha} + (\mu\leftrightarrow\nu) =0. 
\ee
Using  (\ref{CDC}) and noting again that $I$ is Killing, we find that above equation is also trivially satisfied.
 
 \paragraph{$B$-field equation}

We now arrive at a non-trivial equation. We expand the $B$-field equation (\ref{B_eom}), or equivalently (\ref{Beqnew}), and keep the cubic terms in $\Theta$. The first term in  (\ref{Beqnew}) gives
  \bea
  \label{DH}
 &&\frac12 \hat\nabla^\rho H_{\rho\mu\nu}=\frac12g^{\rho\beta} \hat\nabla_\beta H_{\rho\mu\nu}=\frac12(G^{\rho\beta}-\Theta^{\rho\sigma}\Theta_{\sigma\gamma}G^{\gamma\beta} )\hat\nabla_\beta H_{\rho\mu\nu}. 
  \eea
  Using (\ref{CDC}), the covariant derivative can be expanded as
  \be
 \hat\nabla_\beta H_{\rho\mu\nu}=\nabla_\beta H_{\rho\mu\nu}^{(1)}+\nabla_\beta H_{\rho\mu\nu}^{(3)} -C^{\alpha}_{\beta\rho}H_{\alpha\mu\nu}^{(1)}
 -C^{\alpha}_{\beta\mu}H_{\rho\alpha\nu}^{(1)} -C^{\alpha}_{\beta\nu}H_{\rho\mu\alpha}^{(1)} +\mathcal{O}(\Theta^5), 
  \ee
 where $H^{(1)}$  and $H^{(3)}$  are defined in (\ref{H1H3}). Therefore the cubic terms in $\Theta$ in (\ref{DH}) are
  \bea
 \frac12 \nabla^\rho H^{(3)}_{\rho\mu\nu}-\frac12 \Theta^{\rho\sigma}\Theta_\sigma^{\;\;\beta}\nabla_\beta H^{(1)}_{\rho\mu\nu}
 -\frac12 G^{\rho\beta}(C^\alpha_{\beta\rho}H^{(1)}_{\alpha\mu\nu}+C^\alpha_{\beta\mu}H^{(1)}_{\rho\alpha\nu}+C^\alpha_{\beta\nu}H^{(1)}_{\alpha\mu\alpha}). 
  \eea

The second term in equation (\ref{Beqnew}) takes the form
  \be\label{phiHHH}
 -\partial^\rho\phi H_{\rho\mu\nu}^{(3)}=  -\partial^\rho\phi\left[\frac12\nabla_\rho\Theta^3_{\mu\nu}+ \Theta_{\nu\alpha}\Theta^{\alpha\beta}\nabla_\mu\Theta_{\beta\rho}  -(\mu\leftrightarrow \nu)\right], 
 \ee
 where to write down the above equations we have used (\ref{LDphi}). Further, using the following identities:
 \be
\partial^\rho\phi\nabla_\rho \Theta_{\alpha\beta}=\Theta_{\alpha\rho}\nabla_{\beta}\nabla_{\rho}\phi+\Theta_{\rho\beta}\nabla_{\alpha}\nabla_{\rho}\phi,\qquad \partial^\rho\phi\nabla_\mu \Theta_{\alpha\rho}=-\Theta_{\alpha\rho}\nabla^\mu\nabla_\rho \phi, 
 \ee
we can simplify  (\ref{phiHHH}) as follows 
\be
 -\partial^\rho\phi H_{\rho\mu\nu}^{(3)}=\frac12(\Theta^3_{\mu\rho} R_{\nu}^\rho -\Theta_{\nu\rho}^3 R^\rho_\mu). 
\ee
 
 The third term in  (\ref{Beqnew})  takes the form
 \be
B^\rho_{\;\;\;\alpha}I^{\alpha} H_{\rho\mu\nu}=\Theta_\alpha^{~\rho} (\nabla_\rho\Theta_{\mu\nu} + \nabla_\mu\Theta_{\nu\rho}  + \nabla_\nu\Theta_{\rho\mu}) \nabla_\beta\Theta^{\beta\alpha}
 \ee
 
 Adding all the above terms,  massaging them and using bi-Killing structure of $\Theta$, the $B$-field equation at third order can be written as
 \be
\Theta^{\alpha\beta} K_{[\mu}^i K_{\nu}^k \nabla_{\alpha} K_{\beta]}^m \left( c^{l_1 l_2}_{~~~ m} r_{i l_1} r_{k l_2} + c^{l_1 l_2}_{~~~ k} r_{m l_1} r_{i l_2} + c^{l_1 l_2}_{~~~ i} r_{k l_1} r_{m l_2} \right)=0, 
 \ee
where $[\cdots]$ denotes anti-symmetrization with respect to all indices. The above equation is satisfied identically once the CYBE holds.


\begin{thebibliography}{99}

\bibitem{Bakhmatov:2017joy} 
  I.~Bakhmatov, \"O.~Kelekci, E. \'O~Colg\'ain and M.~M.~Sheikh-Jabbari,
  ``Classical Yang-Baxter Equation from Supergravity,''
  arXiv:1710.06784 [hep-th].

\bibitem{Klimcik:2002zj} 
  C.~Klimcik,
  ``Yang-Baxter sigma models and dS/AdS T duality,''
  JHEP {\bf 0212}, 051 (2002)
  [hep-th/0210095].
  
\bibitem{Klimcik:2008eq} 
  C.~Klimcik,
  ``On integrability of the Yang-Baxter sigma-model,''
  J.\ Math.\ Phys.\  {\bf 50}, 043508 (2009)
  [arXiv:0802.3518 [hep-th]].

\bibitem{Delduc:2013qra} 
  F.~Delduc, M.~Magro and B.~Vicedo,
  ``An integrable deformation of the $AdS_5 x S^5$ superstring action,''
  Phys.\ Rev.\ Lett.\  {\bf 112}, no. 5, 051601 (2014)
  [arXiv:1309.5850 [hep-th]].
  
\bibitem{Kawaguchi:2014qwa} 
  I.~Kawaguchi, T.~Matsumoto and K.~Yoshida,
  ``Jordanian deformations of the $AdS_5 x S^5$ superstring,''
  JHEP {\bf 1404}, 153 (2014)
  [arXiv:1401.4855 [hep-th]].

\bibitem{Hashimoto:1999ut}
A.~Hashimoto and N.~Itzhaki, ``Noncommutative Yang-Mills and the AdS/CFT
  correspondence,''
  {Phys. Lett.}
  {\bfseries B465} (1999) 142--147,
[hep-th/9907166]

\bibitem{Maldacena:1999mh}
J.~M. Maldacena and J.~G. Russo, ``{Large $N$ limit of noncommutative gauge
  theories}, {
  JHEP} {\bfseries 09} (1999) 025,
[hep-th/9908134].

\bibitem{Alishahiha:1999ci}
M.~Alishahiha, Y.~Oz, and M.~M. Sheikh-Jabbari, ``{Supergravity and large $N$
  noncommutative field theories},''
  {JHEP} {\bfseries 11} (1999) 007,
[hep-th/9909215]


\bibitem{Lunin:2005jy} 
  O.~Lunin and J.~M.~Maldacena,
  ``Deforming field theories with U(1) x U(1) global symmetry and their gravity duals,''
  JHEP {\bf 0505}, 033 (2005)
  [hep-th/0502086].

\bibitem{Frolov:2005dj} 
  S.~Frolov,
  ``Lax pair for strings in Lunin-Maldacena background,''
  JHEP {\bf 0505}, 069 (2005)
  [hep-th/0503201].

\bibitem{Arutyunov:2015mqj} 
  G.~Arutyunov, S.~Frolov, B.~Hoare, R.~Roiban and A.~A.~Tseytlin,
  ``Scale invariance of the $\eta$-deformed $AdS_5\times S^5$ superstring, T-duality and modified type II equations,''
  Nucl.\ Phys.\ B {\bf 903}, 262 (2016)
  [arXiv:1511.05795 [hep-th]].

\bibitem{Wulff:2016tju} 
  L.~Wulff and A.~A.~Tseytlin,
  ``Kappa-symmetry of superstring sigma model and generalized 10d supergravity equations,''
  JHEP {\bf 1606}, 174 (2016)
  [arXiv:1605.04884 [hep-th]].

\bibitem{Matsumoto:2014nra}
T.~Matsumoto and K.~Yoshida, ``{Lunin-Maldacena backgrounds from the classical
  Yang-Baxter equation - towards the gravity/CYBE correspondence},''
  {{JHEP} {\bfseries 06}
  (2014) 135},
[arXiv:1404.1838 [hep-th]].

\bibitem{Matsumoto:2014gwa}
T.~Matsumoto and K.~Yoshida, ``{Integrability of classical strings dual for
  noncommutative gauge theories},''
  {JHEP} {\bfseries 06}
  (2014) 163,
[arXiv:1404.3657 [hep-th]].

\bibitem{Osten:2016dvf}
D.~Osten and S.~J. van Tongeren, ``{Abelian Yang-Baxter deformations and TsT
  transformations},''
  {Nucl. Phys.}
  {\bfseries B915} (2017) 184--205,
[arXiv:1608.08504 [hep-th]].

\bibitem{Hoare:2016wsk}
B.~Hoare and A.~A. Tseytlin, ``{Homogeneous Yang-Baxter deformations as
  non-abelian duals of the AdS$_5$ sigma-model},''
  { J. Phys.}
  {\bfseries A49} no.~49, (2016) 494001,
[arXiv:1609.02550 [hep-th]].

\bibitem{Borsato:2016pas}
R.~Borsato and L.~Wulff, ``{Integrable Deformations of $T$-Dual $\sigma$
  Models},'' { Phys.
  Rev. Lett.} {\bfseries 117} no.~25, (2016) 251602, [arXiv:1609.09834 [hep-th]].

\bibitem{Borsato:2017qsx} 
  R.~Borsato and L.~Wulff,
  ``On non-abelian T-duality and deformations of supercoset string sigma-models,''
  JHEP {\bf 1710}, 024 (2017)
  [arXiv:1706.10169 [hep-th]].
  
\bibitem{Arutyunov:2013ega}
  G.~Arutyunov, R.~Borsato and S.~Frolov,
  ``S-matrix for strings on $\eta$-deformed AdS5 x S5,''
  JHEP {\bf 1404} (2014) 002
  [arXiv:1312.3542 [hep-th]].
  
\bibitem{Kawaguchi:2014fca}
  I.~Kawaguchi, T.~Matsumoto and K.~Yoshida,
  ``A Jordanian deformation of AdS space in type IIB supergravity,''
  JHEP {\bf 1406} (2014) 146
  [arXiv:1402.6147 [hep-th]].
  
\bibitem{Hoare:2014pna}
  B.~Hoare, R.~Roiban and A.~A.~Tseytlin,
  ``On deformations of $AdS_n$ x $S^n$ supercosets,''
  JHEP {\bf 1406} (2014) 002
  [arXiv:1403.5517 [hep-th]].
  
\bibitem{Arutynov:2014ota}
  G.~Arutyunov, M.~de Leeuw and S.~J.~van Tongeren,
  ``The exact spectrum and mirror duality of the $(\text{AdS}_5{\times}S^5)_\eta$ superstring,''
  Theor.\ Math.\ Phys.\  {\bf 182} (2015) no.1,  23
   [Teor.\ Mat.\ Fiz.\  {\bf 182} (2014) no.1,  28]
  [arXiv:1403.6104 [hep-th]].

\bibitem{Khouchen:2014kaa} 
  M.~Khouchen and J.~Kluson,
  ``Giant Magnon on Deformed AdS(3)xS(3),''
  Phys.\ Rev.\ D {\bf 90}, no. 6, 066001 (2014)
  [arXiv:1405.5017 [hep-th]].
    
\bibitem{Ahn:2014aqa}
  C.~Ahn and P.~Bozhilov,
  ``Finite-size giant magnons on $\eta$-deformed $AdS_5 \times S^5$,''
  Phys.\ Lett.\ B {\bf 737} (2014) 293
  [arXiv:1406.0628 [hep-th]].

\bibitem{Arutyunov:2014cra}
  G.~Arutyunov and S.~J.~van Tongeren,
  ``$\mathrm{AdS}_5 \times \mathrm{S}^5$ mirror model as a string sigma model,''
  Phys.\ Rev.\ Lett.\  {\bf 113} (2014) 261605
  [arXiv:1406.2304 [hep-th]].
  
\bibitem{Arutyunov:2014cda}
  G.~Arutyunov and D.~Medina-Rincon,
  ``Deformed Neumann model from spinning strings on ($AdS_5 \times S^5$)$_\eta$,''
  JHEP {\bf 1410} (2014) 050
  [arXiv:1406.2536 [hep-th]].

\bibitem{Banerjee:2014bca} 
  A.~Banerjee and K.~L.~Panigrahi,
  ``On the rotating and oscillating strings in (AdS$_{3}$  x S$^{3}$)$_{\kappa}$,''
  JHEP {\bf 1409}, 048 (2014)
  [arXiv:1406.3642 [hep-th]].
    
\bibitem{Kameyama:2014vma}
  T.~Kameyama and K.~Yoshida,
  ``A new coordinate system for $q$-deformed AdS$_{5} \times$ S$^5$ and classical string solutions,''
  J.\ Phys.\ A {\bf 48} (2015) no.7,  075401
  [arXiv:1408.2189 [hep-th]].

\bibitem{Kameyama:2014via}
  T.~Kameyama and K.~Yoshida,
  ``Minimal surfaces in $q$-deformed AdS$_5\times$S$^5$ with Poincare coordinates,''
  J.\ Phys.\ A {\bf 48} (2015) no.24,  245401
  [arXiv:1410.5544 [hep-th]].

\bibitem{Lunin:2014tsa}
  O.~Lunin, R.~Roiban and A.~A.~Tseytlin,
  ``Supergravity backgrounds for deformations of AdS$_{n} \times S^n$ supercoset string models,''
  Nucl.\ Phys.\ B {\bf 891} (2015) 106
  [arXiv:1411.1066 [hep-th]].

\bibitem{Hoare:2014oua}
  B.~Hoare,
  ``Towards a two-parameter q-deformation of AdS$_3 \times S^3 \times M^4$ superstrings,''
  Nucl.\ Phys.\ B {\bf 891} (2015) 259
  [arXiv:1411.1266 [hep-th]].

\bibitem{Matsumoto:2014ubv}
  T.~Matsumoto and K.~Yoshida,
  ``Yang-Baxter deformations and string dualities,''
  JHEP {\bf 1503} (2015) 137
  [arXiv:1412.3658 [hep-th]].

\bibitem{Engelund:2014pla} 
  O.~T.~Engelund and R.~Roiban,
  ``On the asymptotic states and the quantum S matrix of the $\eta$-deformed AdS$_{5} \times$ S$^{5}$ superstring,''
  JHEP {\bf 1503}, 168 (2015)
  [arXiv:1412.5256 [hep-th]].
  
\bibitem{Ahn:2014iia}
  C.~Ahn and P.~Bozhilov,
  ``A HHL 3-point correlation function in the $\eta$-deformed AdS$_{5} \times S^5$,''
  Phys.\ Lett.\ B {\bf 743} (2015) 121
  [arXiv:1412.6668 [hep-th]].

\bibitem{Panigrahi:2014sia} 
  K.~L.~Panigrahi, P.~M.~Pradhan and M.~Samal,
  ``Pulsating strings on (AdS$_{3}$ x S$^{3}$)$_{?}$,''
  JHEP {\bf 1503}, 010 (2015)
  [arXiv:1412.6936 [hep-th]].

\bibitem{Bai:2014pya} 
  N.~Bai, H.~H.~Chen and J.~B.~Wu,
  ``Holographic cusped Wilson loops in q-deformed AdS(5) x S(5) spacetime,''
  Chin.\ Phys.\ C {\bf 39}, no. 10, 103102 (2015)
  [arXiv:1412.8156 [hep-th]].
    
\bibitem{Matsumoto:2015jja}
  T.~Matsumoto and K.~Yoshida,
  ``Yang-Baxter sigma models based on the CYBE,''
  Nucl.\ Phys.\ B {\bf 893} (2015) 287
  [arXiv:1501.03665 [hep-th]].

\bibitem{Bozhilov:2015kya} 
  P.~Bozhilov,
  ``Some three-point correlation functions in the $\eta$-deformed AdS$_5 \times S^5$,''
  Int.\ J.\ Mod.\ Phys.\ A {\bf 31}, no. 01, 1550224 (2016)
  [arXiv:1502.00610 [hep-th]].
  
\bibitem{Matsumoto:2015uja}
  T.~Matsumoto and K.~Yoshida,
  ``Schr\"odinger geometries arising from Yang-Baxter deformations,''
  JHEP {\bf 1504} (2015) 180
  [arXiv:1502.00740 [hep-th]].

\bibitem{Banerjee:2015nha} 
  A.~Banerjee, S.~Bhattacharya and K.~L.~Panigrahi,
  ``Spiky strings in $\varkappa$-deformed $AdS$,''
  JHEP {\bf 1506}, 057 (2015)
  [arXiv:1503.07447 [hep-th]].

\bibitem{vanTongeren:2015soa}
  S.~J.~van Tongeren,
  ``On classical Yang-Baxter based deformations of the AdS$_{5}$ x S$^{5}$ superstring,''
  JHEP {\bf 1506} (2015) 048
  [arXiv:1504.05516 [hep-th]].

\bibitem{Vicedo:2015pna} 
  B.~Vicedo,
  ``Deformed integrable $\sigma$-models, classical R-matrices and classical exchange algebra on Drinfeld doubles,''
  J.\ Phys.\ A {\bf 48}, no. 35, 355203 (2015)
  [arXiv:1504.06303 [hep-th]].
  
\bibitem{Hoare:2015gda}
  B.~Hoare and A.~A.~Tseytlin,
  ``On integrable deformations of superstring sigma models related to $AdS_n \times S^n$ supercosets,''
  Nucl.\ Phys.\ B {\bf 897} (2015) 448
  [arXiv:1504.07213 [hep-th]].

\bibitem{Khouchen:2015jfa} 
  M.~Khouchen and J.~Kluso?,
  ``D-brane on deformed AdS$_{3} \times$ S$^{3}$,''
  JHEP {\bf 1508}, 046 (2015)
  [arXiv:1505.04946 [hep-th]].
  
\bibitem{vanTongeren:2015uha}
  S.~J.~van Tongeren,
  ``Yang-Baxter deformations, AdS/CFT, and twist-noncommutative gauge theory,''
  Nucl.\ Phys.\ B {\bf 904} (2016) 148
  [arXiv:1506.01023 [hep-th]].

\bibitem{Sfetsos:2015nya} 
  K.~Sfetsos, K.~Siampos and D.~C.~Thompson,
  ``Generalised integrable $\lambda$- and $\eta$-deformations and their relation,''
  Nucl.\ Phys.\ B {\bf 899}, 489 (2015)
  [arXiv:1506.05784 [hep-th]].

\bibitem{Arutyunov:2015qva}
  G.~Arutyunov, R.~Borsato and S.~Frolov,
  ``Puzzles of $\eta$-deformed AdS$_5 \times$ S$^5$,''
  JHEP {\bf 1512} (2015) 049
  [arXiv:1507.04239 [hep-th]].

\bibitem{Hoare:2015wia}
  B.~Hoare and A.~A.~Tseytlin,
  ``Type IIB supergravity solution for the T-dual of the $\eta$-deformed AdS$_{5} \times$ S$^{5}$ superstring,''
  JHEP {\bf 1510} (2015) 060
  [arXiv:1508.01150 [hep-th]].

\bibitem{Klimcik:2015gba} 
  C.~Klimcik,
  ``$\eta$ and $\lambda$ deformations as E -models,''
  Nucl.\ Phys.\ B {\bf 900}, 259 (2015)
  [arXiv:1508.05832 [hep-th]].
  
\bibitem{Kameyama:2015ufa}
  T.~Kameyama, H.~Kyono, J.~i.~Sakamoto and K.~Yoshida,
  ``Lax pairs on Yang-Baxter deformed backgrounds,''
  JHEP {\bf 1511} (2015) 043
  [arXiv:1509.00173 [hep-th]].

\bibitem{Borowiec:2015wua}
  A.~Borowiec, H.~Kyono, J.~Lukierski, J.~i.~Sakamoto and K.~Yoshida,
  ``Yang-Baxter sigma models and Lax pairs arising from $\kappa$-Poincar\'e $r$-matrices,''
  JHEP {\bf 1604} (2016) 079
  [arXiv:1510.03083 [hep-th]].

\bibitem{Demulder:2016mja}
  S.~Demulder, D.~Dorigoni and D.~C.~Thompson,
  ``Resurgence in $\eta$-deformed Principal Chiral Models,''
  JHEP {\bf 1607} (2016) 088
  [arXiv:1604.07851 [hep-th]].

\bibitem{Kyono:2016jqy}
  H.~Kyono and K.~Yoshida,
  ``Supercoset construction of Yang-Baxter deformed AdS$_5\times$S$^5$ backgrounds,''
  PTEP {\bf 2016} (2016) no.8,  083B03
  [arXiv:1605.02519 [hep-th]].

\bibitem{Hoare:2016ibq}
  B.~Hoare and S.~J.~van Tongeren,
  ``Non-split and split deformations of ${\mathrm{AdS}}_{5}$,''
  J.\ Phys.\ A {\bf 49} (2016) no.48,  484003
  [arXiv:1605.03552 [hep-th]].

\bibitem{Hoare:2016hwh}
  B.~Hoare and S.~J.~van Tongeren,
  ``On jordanian deformations of AdS$_5$ and supergravity,''
  J.\ Phys.\ A {\bf 49} (2016) no.43,  434006
  [arXiv:1605.03554 [hep-th]].

\bibitem{Delduc:2016ihq} 
  F.~Delduc, S.~Lacroix, M.~Magro and B.~Vicedo,
  ``On q-deformed symmetries as Poisson-Lie symmetries and application to Yang-Baxter type models,''
  J.\ Phys.\ A {\bf 49}, no. 41, 415402 (2016)
  [arXiv:1606.01712 [hep-th]].

\bibitem{Klimcik:2016rov} 
  C.~Klimcik,
  ``Poisson-Lie T-duals of the bi-Yang-Baxter models,''
  Phys.\ Lett.\ B {\bf 760}, 345 (2016)
  [arXiv:1606.03016 [hep-th]].
  
\bibitem{Orlando:2016qqu}
  D.~Orlando, S.~Reffert, J.~i.~Sakamoto and K.~Yoshida,
  ``Generalized type IIB supergravity equations and non-Abelian classical r-matrices,''
  J.\ Phys.\ A {\bf 49} (2016) no.44,  445403
  [arXiv:1607.00795 [hep-th]].

\bibitem{Banerjee:2016xbb} 
  A.~Banerjee and K.~L.~Panigrahi,
  ``On circular strings in $(AdS_3 \times S^3)_{\varkappa}$,''
  JHEP {\bf 1609}, 061 (2016)
  [arXiv:1607.04208 [hep-th]].
  
\bibitem{Arutyunov:2016ysi}
  G.~Arutyunov, M.~Heinze and D.~Medina-Rincon,
  ``Integrability of the $\eta$-deformed Neumann-Rosochatius model,''
  J.\ Phys.\ A {\bf 50} (2017) no.3,  035401
  [arXiv:1607.05190 [hep-th]].

\bibitem{vanTongeren:2016eeb}
  S.~J.~van Tongeren,
  ``Almost abelian twists and AdS/CFT,''
  Phys.\ Lett.\ B {\bf 765} (2017) 344
  [arXiv:1610.05677 [hep-th]].

\bibitem{Hoare:2016wca}
  B.~Hoare and D.~C.~Thompson,
  ``Marginal and non-commutative deformations via non-abelian T-duality,''
  JHEP {\bf 1702} (2017) 059
  [arXiv:1611.08020 [hep-th]].

\bibitem{Ahn:2016egk}
  C.~Ahn,
  ``Finite-size effect of $\eta$-deformed AdS$_5 \times$ S$^5$ at strong coupling,''
  Phys.\ Lett.\ B {\bf 767} (2017) 121
  [arXiv:1611.09992 [hep-th]].

\bibitem{Roychowdhury:2016bsv} 
  D.~Roychowdhury,
  ``Multispin magnons on deformed $ AdS_{3}\times S^{3} $,''
  Phys.\ Rev.\ D {\bf 95}, no. 8, 086009 (2017)
  [arXiv:1612.06217 [hep-th]].
  
\bibitem{Sakamoto:2016ppx}
  J.~i.~Sakamoto and K.~Yoshida,
  ``Yang-Baxter deformations of $W_{2,4}\times T^{1,1}$ and the associated T-dual models,''
  Nucl.\ Phys.\ B {\bf 921} (2017) 805
  [arXiv:1612.08615 [hep-th]].

\bibitem{Delduc:2017brb}
  F.~Delduc, T.~Kameyama, M.~Magro and B.~Vicedo,
  ``Affine $q$-deformed symmetry and the classical Yang-Baxter  $\sigma$-model,''
  JHEP {\bf 1703} (2017) 126
  [arXiv:1701.03691 [hep-th]].
  
\bibitem{Kyono:2017jtc}
  H.~Kyono, S.~Okumura and K.~Yoshida,
  ``Deformations of the Almheiri-Polchinski model,''
  JHEP {\bf 1703} (2017) 173
  [arXiv:1701.06340 [hep-th]].

\bibitem{Roychowdhury:2017oqd} 
  D.~Roychowdhury,
  ``Stringy correlations on deformed $ AdS_{3}\times S^{3} $,''
  JHEP {\bf 1703}, 043 (2017)
  [arXiv:1702.01405 [hep-th]].
  
\bibitem{Appadu:2017bnv}
  C.~Appadu, T.~J.~Hollowood, D.~Price and D.~C.~Thompson,
  ``Yang Baxter and Anisotropic Sigma and Lambda Models, Cyclic RG and Exact S-Matrices,''
  JHEP {\bf 1709} (2017) 035
  [arXiv:1706.05322 [hep-th]].

\bibitem{Klimcik:2017ken}
  C.~Klimcik,
  ``Yang-Baxter $\sigma$-model with WZNW term as ${ \mathcal E}$-model,''
  Phys.\ Lett.\ B {\bf 772} (2017) 725
  [arXiv:1706.08912 [hep-th]].

\bibitem{Roychowdhury:2017vdo} 
  D.~Roychowdhury,
  ``Analytic integrability for strings on $ \eta $ and $ \lambda $ deformed backgrounds,''
  JHEP {\bf 1710}, 056 (2017)
  [arXiv:1707.07172 [hep-th]].

\bibitem{Hernandez:2017raj} 
  R.~Hernandez and J.~M.~Nieto,
  ``Spinning strings in the $\eta$-deformed Neumann-Rosochatius system,''
  Phys.\ Rev.\ D {\bf 96}, no. 8, 086010 (2017)
  [arXiv:1707.08032 [hep-th]].
    
\bibitem{Delduc:2017fib}
  F.~Delduc, B.~Hoare, T.~Kameyama and M.~Magro,
  ``Combining the bi-Yang-Baxter deformation, the Wess-Zumino term and TsT transformations in one integrable $\sigma$-model,''
  JHEP {\bf 1710} (2017) 212
  [arXiv:1707.08371 [hep-th]].

\bibitem{Klabbers:2017vtw}
  R.~Klabbers and S.~J.~van Tongeren,
  ``Quantum Spectral Curve for the eta-deformed AdS$_5$xS$^5$ superstring,''
  Nucl.\ Phys.\ B {\bf 925} (2017) 252
  [arXiv:1708.02894 [hep-th]].

\bibitem{Hoare:2017ukq}
  B.~Hoare and F.~K.~Seibold,
  ``Poisson-Lie duals of the $\eta$ deformed symmetric space sigma model,''
  JHEP {\bf 1711} (2017) 014
  [arXiv:1709.01448 [hep-th]].

\bibitem{Demulder:2017zhz}
  S.~Demulder, S.~Driezen, A.~Sevrin and D.~C.~Thompson,
  ``Classical and Quantum Aspects of Yang-Baxter Wess-Zumino Models,''
  JHEP {\bf 1803} (2018) 041
  [arXiv:1711.00084 [hep-th]].

\bibitem{Banerjee:2017mpe} 
  A.~Banerjee, A.~Bhattacharyya and D.~Roychowdhury,
  ``Fast spinning strings on $ \eta $ deformed $ AdS_5 \times S^{5} $,''
  JHEP {\bf 1802}, 035 (2018)
  [arXiv:1711.07963 [hep-th]].

\bibitem{Barik:2018haz} 
  S.~P.~Barik, K.~L.~Panigrahi and M.~Samal,
  ``Spinning pulsating strings in $(AdS_5 \times S^5)_{\varkappa}$,''
  arXiv:1801.04248 [hep-th].
    
\bibitem{Lust:2018jsx}
  D.~Lust and D.~Osten,
  ``Generalised fluxes, Yang-Baxter deformations and the O(d,d) structure of non-abelian T-duality,''
  arXiv:1803.03971 [hep-th].
  
\bibitem{Delduc:2014kha} 
  F.~Delduc, M.~Magro and B.~Vicedo,
  ``Derivation of the action and symmetries of the $q$-deformed $AdS_{5} \times S^{5}$ superstring,''
  JHEP {\bf 1410}, 132 (2014)
  [arXiv:1406.6286 [hep-th]].

\bibitem{Elitzur:1994ri} 
  S.~Elitzur, A.~Giveon, E.~Rabinovici, A.~Schwimmer and G.~Veneziano,
  ``Remarks on nonAbelian duality,''
  Nucl.\ Phys.\ B {\bf 435}, 147 (1995)
  [hep-th/9409011].
  

\bibitem{Hull-Tonwsend} 
  C.~M.~Hull and P.~K.~Townsend,
  ``String Effective Actions From $\sigma$ Model Conformal Anomalies,''
  Nucl.\ Phys.\ B {\bf 301}, 197 (1988).
          
\bibitem{Araujo:2017jkb}
  T.~Araujo, I.~Bakhmatov, E. \'O~Colg\'ain, J.~Sakamoto, M.~M.~Sheikh-Jabbari and K.~Yoshida,
  ``Yang-Baxter $\sigma$-models, conformal twists, and noncommutative Yang-Mills theory,''
  Phys.\ Rev.\ D {\bf 95}, no. 10, 105006 (2017)
  [arXiv:1702.02861 [hep-th]].

\bibitem{Araujo:2017jap} 
  T.~Araujo, I.~Bakhmatov, E. \'O~Colg\'ain, J.~i.~Sakamoto, M.~M.~Sheikh-Jabbari and K.~Yoshida,
  ``Conformal Twists, Yang-Baxter $\sigma$-models \& Holographic Noncommutativity,''
  arXiv:1705.02063 [hep-th].
    
\bibitem{Araujo:2017enj} 
  T.~Araujo, E. \'O~Colg\'ain, J.~Sakamoto, M.~M.~Sheikh-Jabbari and K.~Yoshida,
  ``$I$ in generalized supergravity,''
  Eur.\ Phys.\ J.\ C {\bf 77}, no. 11, 739 (2017)
  [arXiv:1708.03163 [hep-th]].

\bibitem{Seiberg:1999vs} 
  N.~Seiberg and E.~Witten,
  ``String theory and noncommutative geometry,''
  JHEP {\bf 9909}, 032 (1999)
  [hep-th/9908142].

\bibitem{Duff:1989tf} 
  M.~J.~Duff,
  ``Duality Rotations in String Theory,''
  Nucl.\ Phys.\ B {\bf 335}, 610 (1990).


\bibitem{Sakamoto:2017cpu} 
  J.~i.~Sakamoto, Y.~Sakatani and K.~Yoshida,
  ``Homogeneous Yang-Baxter deformations as generalized diffeomorphisms,''
  J.\ Phys.\ A {\bf 50}, no. 41, 415401 (2017)
  [arXiv:1705.07116 [hep-th]].
  
\bibitem{Fernandez-Melgarejo:2017oyu}
  J.~J.~Fernandez-Melgarejo, J.~i.~Sakamoto, Y.~Sakatani and K.~Yoshida,
  ``$T$-folds from Yang-Baxter deformations,''
  JHEP {\bf 1712} (2017) 108
  [arXiv:1710.06849 [hep-th]].

\bibitem{Sakamoto:2018krs}
  J.~i.~Sakamoto and Y.~Sakatani,
  ``Local $\beta$-deformations and Yang-Baxter sigma model,''
  arXiv:1803.05903 [hep-th].

\bibitem{Crichigno:2014ipa} 
  P.~M.~Crichigno, T.~Matsumoto and K.~Yoshida,
  ``Deformations of $T^{1,1}$ as Yang-Baxter sigma models,''
  JHEP {\bf 1412}, 085 (2014)
  [arXiv:1406.2249 [hep-th]].
  
\bibitem{Hassan:1999bv} 
  S.~F.~Hassan,
  ``T duality, space-time spinors and RR fields in curved backgrounds,''
  Nucl.\ Phys.\ B {\bf 568}, 145 (2000)
  [hep-th/9907152].

\bibitem{Kelekci:2014ima} 
  \"O.~Kelekci, Y.~Lozano, N.~T.~Macpherson and E.~\'O~Colg\'ain,
  ``Supersymmetry and non-Abelian T-duality in type II supergravity,''
  Class.\ Quant.\ Grav.\  {\bf 32}, no. 3, 035014 (2015)
  [arXiv:1409.7406 [hep-th]].
        
\bibitem{Borsato:2016ose} 
  R.~Borsato and L.~Wulff,
  ``Target space supergeometry of $\eta$ and $\lambda$-deformed strings,''
  JHEP {\bf 1610}, 045 (2016)
  [arXiv:1608.03570 [hep-th]].

\bibitem{Drinfeld}
V.~G. Drinfeld {\em Leningrad Math. J.} {\bfseries 1} (1990) 321.

\bibitem{Chaichian:2004za}
M.~Chaichian, P.~P. Kulish, K.~Nishijima, and A.~Tureanu, ``{On a
  Lorentz-invariant interpretation of noncommutative space-time and its
  implications on noncommutative QFT},''
  {Phys. Lett.}
  {\bfseries B604} (2004) 98--102,
[hep-th/0408069]

\bibitem{Chaichian:2004yh}
M.~Chaichian, P.~Presnajder, and A.~Tureanu, ``{New concept of relativistic
  invariance in NC space-time: Twisted Poincare symmetry and its
  implications},'' {
  Phys. Rev. Lett.} {\bfseries 94} (2005) 151602,
[hep-th/0409096]

\bibitem{Lukierski:2005fc}
J.~Lukierski and M.~Woronowicz, ``{New Lie-algebraic and quadratic deformations
  of Minkowski space from twisted Poincare symmetries},''
 { Phys. Lett.}
  {\bfseries B633} (2006) 116--124,
[arXiv:hep-th/0508083
  [hep-th]].

\bibitem{Hong:2018tlp} 
  M.~Hong, Y.~Kim and E. \'O~Colg\'ain,
  ``On non-Abelian T-duality for non-semisimple groups,''
  arXiv:1801.09567 [hep-th].
    
\bibitem{Matsumoto:2015ypa} 
  T.~Matsumoto, D.~Orlando, S.~Reffert, J.~i.~Sakamoto and K.~Yoshida,
  ``Yang-Baxter deformations of Minkowski spacetime,''
  JHEP {\bf 1510}, 185 (2015)
  [arXiv:1505.04553 [hep-th]].

\bibitem{Page:1984qv} 
  D.~N.~Page,
  ``Classical Stability of Round and Squashed Seven Spheres in Eleven-dimensional Supergravity,''
  Phys.\ Rev.\ D {\bf 28}, 2976 (1983).
         
\bibitem{Rennecke:2014sca} 
  F.~Rennecke,
  ``O(d,d)-Duality in String Theory,''
  JHEP {\bf 1410}, 69 (2014)
  [arXiv:1404.0912 [hep-th]].
              
\end{thebibliography}
\end{document}